\documentclass[12pt,preprint]{elsarticle}

\usepackage{lineno,hyperref,amssymb,graphicx,float,subcaption}
\usepackage{amsmath,bm, amsthm}
\usepackage{color}
\usepackage{geometry}
\usepackage{pdflscape}
\modulolinenumbers[5]
\graphicspath{{figures/}}
\theoremstyle{remark}
\newdefinition{remark}{Remark}

\journal{Journal of Computational Physics}

\newcommand{\etal}{\textit{et al}.}

\newcommand{\norm}[1]{\left\lVert#1\right\rVert}

\begin{document}

\begin{frontmatter}

\title{Quantifying model form uncertainty in Reynolds-averaged turbulence models with Bayesian deep neural networks}

\author[label1]{Nicholas Geneva}
\ead{ngeneva@nd.edu}

\author[label1]{Nicholas Zabaras\corref{cor1}}
\ead{nzabaras@gmail.com}
\ead[url]{https://cics.nd.edu/}

\address[label1]{Center for Informatics and Computational Science, University of Notre Dame, 311 I Cushing Hall, Notre Dame, IN 46556, USA}
\cortext[cor1]{Corresponding author}

\begin{abstract}
Data-driven methods for improving turbulence modeling in Reynolds-Averaged Navier-Stokes (RANS) simulations have gained significant interest in the computational fluid dynamics community. 
Modern machine learning algorithms have opened up a new area of black-box turbulence models allowing for the tuning of RANS simulations to increase their predictive accuracy.
While several data-driven turbulence models have been reported, the quantification of the uncertainties introduced has mostly been neglected.
Uncertainty quantification for such data-driven models is essential since their predictive capability rapidly declines as they are tested for flow physics that deviate from that in the training data.
In this work, we propose a novel data-driven framework that not only improves RANS predictions but also provides probabilistic bounds for fluid quantities such as velocity and pressure. 
The uncertainties capture both model form uncertainty as well as epistemic uncertainty induced by the limited training data.
An invariant Bayesian deep neural network is used  to predict the anisotropic tensor component  of the Reynolds stress.
This model is trained using  Stein variational gradient decent algorithm. The computed uncertainty on the Reynolds stress is propagated to  the quantities of interest by vanilla Monte Carlo simulation.
Results are presented for two test cases  that differ geometrically from the training flows  at several different Reynolds numbers.
The prediction enhancement of the data-driven model is discussed as well as the associated probabilistic bounds for flow properties of interest.
Ultimately this framework allows for a quantitative measurement of model confidence and uncertainty quantification for flows in which no high-fidelity observations or prior knowledge is available.
\end{abstract}


\end{frontmatter}


\section{Introduction}

Over the past decade, with the exponential power increase of computer hardware, computational fluid dynamics (CFD) has become an ever more predominate tool for fluid flow analysis.
The Reynolds-averaged Navier-Stokes (RANS) equation provides an efficient method to compute time-averaged turbulent flow quantities making RANS solvers a frequently selected CFD method. 
However, it is common knowledge that RANS simulations can be highly inaccurate for a variety of flows due to the modeling of the Reynolds stress term~\cite{pope2001turbulent}.
Although over recent years Large Eddy Simulations (LES) or Direct Numerical Simulations (DNS) have become more accessible, these methods still  remain out of the scope of practical engineering applications.
For example, design and optimization tasks require repeated simulations with rapid turnaround time requirements for which RANS simulations are the choice modeling tool.
Thus improving the accuracy of RANS simulations and providing measures of their predictive capability remains essential for the CFD community.

Turbulence models seek to resolve the closure problem that is brought about from the time averaging of the Navier-Stokes equations.
While CFD and computational technology has made significant strides over the past decade, turbulence models have largely become stagnate with the majority of today's most popular models being developed over two decades ago.
Many of the most widely used turbulence models employ the Boussinesq assumption as the theoretical foundation combined with a set of parameters that are described through one or more transport equations.
In general, these turbulence models can be broken down into families based off the number of additional partial differential equations they introduce into the system.
For example, the Spalart-Allmaras model~\cite{spalart1992one} belongs to the family of single equation models.
While the Spalart-Allmaras model has been proven to be useful for several aerodynamic related flows~\cite{godin1997high}, its very general structure severely limits the range of flows that it is applicable.
In the two-equation family, models such as the k-$\epsilon$ model~\cite{jones1972prediction, launder1974application} and the k-$\omega$ model~\cite{wilcox1993turbulence} provide better modeling for a much larger set of flows even though their limitations are well known.
In all the aforementioned models, a set of empirically found constants are used for model-calibration thus resulting in potentially poor performance for flows that were not considered in the calibration process.
This combined with empirical modeling of specific transport equations, such as the $\epsilon$ equation, result in a significant source of model form uncertainty.
While many have proposed more complex approaches such as using different turbulence models for different regions of the flow~\cite{menter1994two} or using a turbulence model with additional transport equations~\cite{walters2008three}, these methods still rely heavily on empirical tuning and calibration.
Thus model form uncertainty introduced by turbulence models continues to be one of the largest sources of uncertainty in RANS simulations.

This work aims to improve turbulence modeling for RANS simulations using machine learning techniques that also allow us to quantify the underlying model error.  
While the use of machine learning methods in CFD simulations can be traced back to over a decade ago~\cite{milano2002neural}, 
recently there has been a new wave of integrating innovative machine learning algorithms to quantify and improve the accuracy of CFD simulations.
Earlier work in quantifying the uncertainty and calibration of turbulence models focused on treating model parameters as random variables and sampling via Monte Carlo to obtain a predictive  distribution of outcomes~\cite{cheung2011bayesian, oliver2011bayesian}.
Rather than constraining oneself to a specific model, an alternative approach was to directly perturb components of the anisotropy term of the Reynolds stress~\cite{dow2011uncertainty}.
Lately, the use of machine learning models has been shown to provide an efficient alternative to direct sampling.
In general, the integration of machine learning with turbulence models can be broken down into three different approaches: modeling the anisotropic term of the Reynolds stress directly, modeling the coefficients of turbulence models and modeling new terms in the turbulence model.
Tracey \etal~\cite{tracey2013application} explored the use of kernel regression to model the eigenvalues of the anisotropic term of the Reynolds stress.
Later, Tracey \etal~\cite{tracey2015machine} used a single layer neural network to predict a source term in the Spalart-Allmaras turbulence model.
Similarly, Signh \etal~\cite{singh2017machine} have used neural networks to introduce a functional corrective term to the source term of the Spalart-Allmaras turbulent model for predicting various quantities over airfoils.
Zhang \etal~\cite{zhang2015machine}  investigated the use of neural networks and Gaussian processes to model a correction term introduced to the turbulence model.
Ling \etal~\cite{ling2016reynolds} considered deep neural networks to predict the anisotropic tensor using a neural network structure with embedded invariance~\cite{ling2016machine}.
Ling \etal~\cite{ling2017uncertainty} additionally proposed using random forests to improve RANS predictions for a flow with a jet in a cross flow.

While the above works have managed to improve the accuracy of RANS simulations,  uncertainty quantification has largely been ignored.
Arguably, the integration of black box machine learning models increases the importance of uncertainty quantification in the context of quantifying the error of the improved turbulence model but also quantifying the uncertainty of the machine learning predictions.
This is largely due to the significant prediction degradation of these proposed machine learning models for flows that  vary from the training data in either fluid properties or geometry~\cite{tracey2013application, ling2016reynolds}.
Past literature has clearly shown that data-driven methods are not exempt from the conflicting objectives of predictive accuracy versus flow versatility seen in traditional turbulence modeling.

Several works have taken steps towards using machine learning to provide uncertainty quantification analysis of RANS simulations.
For example, Xiao \etal~\cite{xiao2016quantifying} proposed a Bayesian data-driven methodology that uses a set of high-fidelity observations to iteratively tune an ensemble of Reynolds-stress fields and other quantities of interest.
While proven to work well for even sparse observational data, this work is limited to a single flow with which the machine learning model was trained explicitly on.
Wu \etal~\cite{wu2017priori} used the Mahalanobis distance and kernel density estimation to formulate a method to predict the confidence of a data-driven model for a given flow.
While this allows the potential identification of regions of less confidence after training, it is limited to the prediction of the anisotropic stress and fails to provide any true probabilistic bounds.

For machine learning methods to be a practical tool for reliably tuning RANS turbulence models, transferability to flows with different geometries and fluid properties is important.
Additionally, quantifying the model uncertainty is critical for assessing both the accuracy and confidence of the machine learning model and of the resulting predicted quantities of interest.

The novelty of our work is  the use of a data-driven model with a Bayesian deep learning framework to provide the means of improving the accuracy of RANS simulations and allow for the quantification of the model form uncertainty arising in the turbulence model.
This uncertainty is then propagated to the quantities of interest, such as pressure and velocity.
The focus of our work will not be application on flows that are the same or similar to those in the training set, but rather to flows defined by different geometries and fluid properties.
We aim to take a much more practical and expansive view of using these innovative machine learning models for improved turbulence modeling.
The specific novel contributions of this work are fourfold: (a) the use of a Bayesian deep neural network as a model to predict a tuned Reynolds stress field, (b) introducing   a stochastic data-driven RANS algorithm that allows us to calculate   probabilistic bounds   for any flow field quantity, (c) assessment of the data-driven model on flows that are geometrically different from the training simulations and (d) comparison of both performance and confidence of the data-driven model across several Reynolds numbers.

This paper is structured as the following:
In Section~\ref{sec:Formulation}, we review the governing equations and motivation for this work. 
In Section~\ref{sec:Framework}, the proposed data-driven framework is discussed in detail.
We discuss the invariant machine learning model in Section~\ref{subsec:invarnn}, its extension to the Bayesian paradigm in Section~\ref{subsec:svgd} and the stochastic data-driven RANS methodology to propagate uncertainty from the Bayesian data-driven model to quantities of interest in Section~\ref{subsec:uq}. 
In Section~\ref{sec:Training}, various implementation details are reviewed including information regarding flow data used, training techniques and integration in the selected CFD solver.
Section~\ref{sec:NumericalResults} details the results of applying this model to two test flows at three different Reynolds numbers.
Results for a flow over a backwards step and over a wall mounted cube are presented in Sections~\ref{subsec:backwardsStep} and~\ref{subsec:wallMountedCube}, respectively.
Finally discussion and conclusions are provided in Section~\ref{sec:Conclusions}.

\section{Problem Formulation}
\label{sec:Formulation}
\subsection{Governing Equations}
\noindent As previously mentioned, the difficulty of RANS is the fundamental closure problem that is introduced when the Navier-Stokes equations are averaged with respect to time.
The RANS momentum equation is as follows:
\begin{equation}
    \left<u_{j}\right>\frac{\partial \left<u_{i}\right>}{\partial x_{j}} = \frac{\partial}{\partial x_{j}}\left[ -\frac{\left<p\right>}{\rho}\delta_{ij} + \nu\left(\frac{\partial \left<u_{i}\right>}{\partial x_{j}} + \frac{\partial \left<u_{j}\right>}{\partial x_{i}} \right) - \left<u'_{i}u'_{j}\right> \right] + \left<g_{i}\right>.\label{eq:rans}
\end{equation}
As always, the challenge is to close this equation by approximating the Reynolds stress (R-S) term $\left< u'_{i}u'_{j}\right>$.
$u'_{i}$ indicates a fluctuation velocity defined as $u_{i}(x,t)-\left< u_{i}(x,t) \right>$ in which $\left< \cdot \right>$ indicates time-averaged or mean value.
The turbulent viscosity theory, originally developed by Boussinesq~\cite{boussinesq1877mem}, proposes a form of the R-S that is mathematically analogous to the stress-strain rate of a Newtonian fluid:
\begin{gather}
    \left<u'_{i}u'_{j}\right> = \frac{2}{3}\delta_{ij}k + a_{ij}, \label{eq:rs-1}\\
    k = \frac{1}{2}\left<u'_{k}u'_{k}\right>, \label{eq:tke}\\
    a_{ij} = -\nu_{t}\left(\frac{\partial\left<u_{i}\right>}{\partial x_{j}}+\frac{\partial \left<u_{j}\right>}{\partial x_{i}}-\frac{2}{3}\delta_{ij}\frac{\partial \left<u_{k}\right>}{\partial x_{k}}\right),
\end{gather}
where $k$, $\nu_{t}$ are the turbulent kinetic energy (TKE) and turbulent viscosity, respectively.
Assuming that the flow is incompressible results in the following:
\begin{equation}
    \left<u'_{i}u'_{j}\right>= \frac{2}{3}\delta_{ij}k - \nu_{t}\left(\frac{\partial\left<u_{i}\right>}{\partial x_{j}}+\frac{\partial \left<u_{j}\right>}{\partial x_{i}}\right).
\end{equation}
This representation is used not because of its accuracy  but instead due to the simplifications that result when it is substituted into the RANS equation.
This form is known as the \textit{Boussinesq eddy viscosity assumption}.

\subsection{RANS Turbulence Models}
\noindent
The context of this work is focused on the $k-\epsilon$ turbulence model~\cite{jones1972prediction, launder1983numerical, chien1982predictions} which is the most commonly used closure model for RANS simulations to date~\cite{pope2001turbulent}.
Starting with the Boussinesq eddy viscosity assumption, the $k-\epsilon$ model approximates the effective viscosity $\nu_{t}$ in terms of the turbulent kinetic energy $k$ and the turbulent dissipation rate $\epsilon$ with the R-S given as follows:
\begin{gather}
    \left<u'_{i}u'_{j}\right> = -\tau_{ij}=  \frac{2}{3}\delta_{ij}k - \nu_{t}\left(\frac{\partial\left<u_{i}\right>}{\partial x_{j}}+\frac{\partial \left<u_{j}\right>}{\partial x_{i}}\right),\\
    \nu_{t}=\frac{C_{\mu} k^{2}}{\epsilon},
\end{gather}
where $C_{\mu}$ is one of five model constants.
Through manipulation of the Navier-Stokes equations,  the kinetic energy can be derived precisely for the case of   high Reynolds number. 
On the other hand, the standard transport equation for the turbulent dissipation, $\epsilon$, should be thought of as an empirical fit~\cite{pope2001turbulent}.
For this work, we will use the standard $k-\epsilon$ model for fully-turbulent, incompressible flow~\cite{wilcox1993turbulence}:
\begin{gather}
    \frac{\partial k}{\partial t}+\left<u_{i}\right>\frac{\partial k}{\partial x_{i}}=\frac{\partial}{\partial x_{i}}\left[\left(\nu+\frac{\nu_{t}}{\sigma_{k}}\right)\frac{\partial k}{\partial x_{i}}\right]+\tau_{ij}\frac{\partial \left<u_{i}\right>}{\partial x_{j}} - \epsilon,\\
    \frac{\partial \epsilon}{\partial t} + \left<u_{i}\right>\frac{\partial\epsilon}{\partial x_{i}} = \frac{\partial}{\partial x_{i}}\left[\left(\nu+\frac{\nu_{t}}{\sigma_{\epsilon}}\right)\frac{\partial \epsilon}{\partial x_{i}}\right] +C_{\epsilon 1}\frac{\epsilon}{k}\tau_{ij}\frac{\partial \left<u_{i}\right>}{\partial x_{j}} - C_{\epsilon 2}\frac{\epsilon^{2}}{k}.
\end{gather}
The five constants $ C_{\mu}, C_{\epsilon 1}, C_{\epsilon 2}, \sigma_{k}, \sigma_{\epsilon}$ are tunable parameters whose optimal values depend on the flow under consideration.
We use the values originally proposed by Launder \etal~\cite{launder1983numerical} obtained by data fitting over various turbulent flows:
\begin{equation}
    C_{\mu}=0.09,\quad C_{\epsilon 1}=1.44, \quad C_{\epsilon 2}=1.92, \quad \sigma_{k}=1.0, \quad \sigma_{\epsilon}=1.3.
\end{equation}

The advantages of the $k-\epsilon$ model are its numerical robustness, computational efficiency, easy implementation and general validity for fully-turbulent flows.
However, with this versatility comes some significant drawbacks including poor accuracy for complex fluid flows, and for problems with flow separation and sharp pressure gradients~\cite{menter1994two, menter1993zonal}.
Core assumptions such as the formulation of the turbulent dissipation equations, the turbulent model constants and even the Boussinesq approximation provide large sources of uncertainty for the $k-\epsilon$ model.
Converged simulations using the $k-\epsilon$ model with the parameters discussed above will be referred to as \textit{baseline} RANS simulations.
Ultimately, we seek to improve the prediction of a baseline simulation through the proposed data-driven framework.

\section{Data-Driven Framework}
\label{sec:Framework}
\begin{figure}[h]
    \centering
    \includegraphics[width=0.8\textwidth]{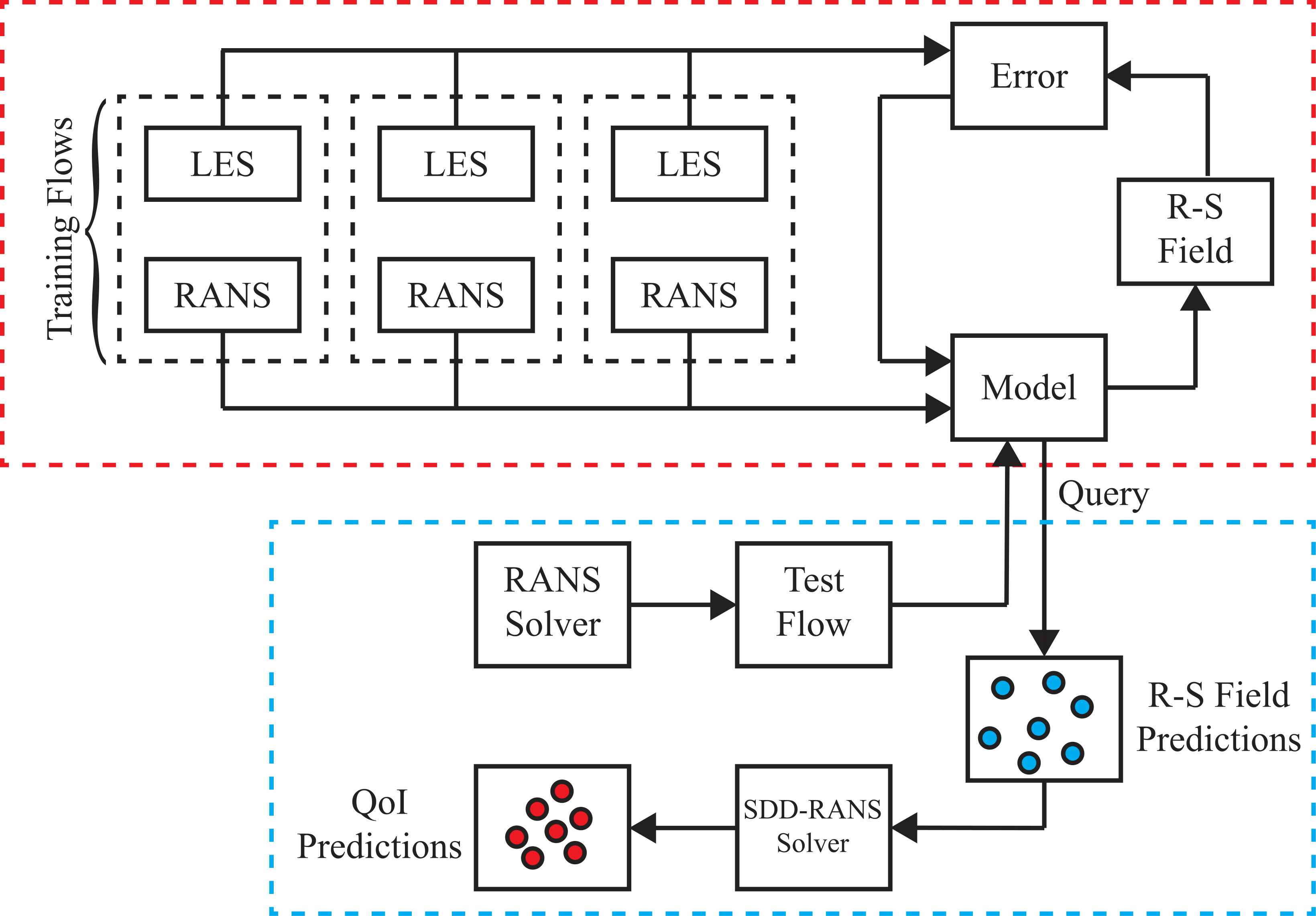}
    \caption{A schematic of the data-driven Bayesian machine learning framework. The top block illustrates the model training using a set of different flows.
    Once trained, the model is then queried given a baseline RANS flow and a set of Reynolds stress (R-S) field realizations are sampled.
    Independent RANS simulations are then performed using these predicted fields by stochastic data-driven RANS (SDD-RANS) and statistics for quantities of interest (QoI) are collected.}
    \label{fig:workflow}
\end{figure}

\noindent
In this work, our goal is to introduce a data-driven model to increase the accuracy of a given RANS simulation and to provide uncertainty bounds for quantities of interest thus capturing the error of the turbulence model.
The proposed framework is illustrated in Fig.~\ref{fig:workflow} which, in a broad sense, shares similar characteristics to earlier works on data-driven turbulence models~\cite{ling2016reynolds, xiao2016quantifying, singh2017machine}.
However, we introduce several novel modifications to the process.
We break this framework down into two key phases: the training of a model using a set of pre-existing flow data and the prediction stage for which the model is sampled to produce fluid flow responses.

The training data that is driving our model is a small library of different fluid flows that attempt to capture different fluid physics.
Ideally each training flow should bring new information for the model to learn thus increasing its potential predictive capability. 
For each unique flow, there is a low-fidelity RANS solution and a time-averaged high-fidelity LES solution.
The objective of this model is to learn the mapping from some baseline RANS flow input information to a turbulent property yielding an improved R-S field matching that of the corresponding high-fidelity simulation.
This turbulent property could be tuned model coefficients, model correction terms or components of the R-S directly.
For the scope of this work, we will focus on modeling the R-S tensor directly but this framework can extend to other approaches.
An error or loss function that quantifies the discrepancy between the predicted R-S and the true high-fidelity field is used to update the model in an iterative process.
We select a Bayesian neural network to serve as this model. Its formulation is discussed in Section~\ref{subsec:invarnn} with a Bayesian extension presented in Section~\ref{subsec:svgd}.
The methods and techniques used to train the model are outlined in Section~\ref{sec:Training}.

Once the model has been trained, it can be used as a regression model to sample predicted R-S fields for a given reference RANS solution.
This process starts with a baseline RANS simulation whose flow field will serve as the input into the calibrated model.
From this model, a set of turbulent properties are sampled that correspond to a predicted high-fidelity representation of the R-S field.
For each predicted field, an independent forward simulation is completed in which the R-S is held constant and the remaining state variables are relaxed around the predicted field from their baseline values to updated perturbed values.
We refer to this process of executing an ensemble of forward simulations as stochastic data-driven RANS (SDD-RANS).
The forward simulations for different samples of the R-S can then be used to compute statistical bounds for quantities of interest as discussed in Section~\ref{subsec:uq}.

\begin{remark}
LES has been chosen as the high-fidelity method for obtaining the training data in the context of this work. 
However, one could alternatively use higher accuracy methods such as DNS or even a combination of methods assuming that their turbulent statistics are consistent.
The use of LES introduces   potential physical inconsistencies  in the high-fidelity predictions.
Namely, LES  can yield different results for the same flow depending on various parameters such as the mesh resolution or the subgrid-scale model used.
The use of more consistent DNS data will likely make training more efficient and increase predictive accuracy.
\end{remark}

\subsection{Invariant Neural Network}
\label{subsec:invarnn}
\noindent
As previously mentioned, in the scope of this work, we will predict the R-S field directly by the  anisotropic component shown in Eq.~\eqref{eq:rs-1}. This approach has been used by multiple earlier works~\cite{ling2016reynolds,xiao2016quantifying}.
These works consider an explicit representation of the R-S component, specifically in the form of a constant field.
The remaining fluid flow quantities (mean velocity and pressure) are then \textit{propagated} forward by solving a numerical system around this constrained R-S field.
Such an approach does not constrain our work to  model-specific assumptions.
We note that explicit R-S approaches can potentially result in significant prediction error of fluid quantities when the Reynolds number approaches $5000$ and above~\cite{thompson2016methodology, wu2018rans}.
Additionally, small errors in the predicted R-S field in an explicit representation can be amplified leading to instabilities.
We accept this as an open problem and while alternative implicit approaches have been  proposed~\cite{wu2018rans}, we leave the discussion of such methods to future works.
However, we will show how the proposed framework can reflect such difficulties through the predicted probabilistic bounds on the flow quantities of interest.

\begin{figure}[H]
    \centering
    \includegraphics[width=0.8\textwidth]{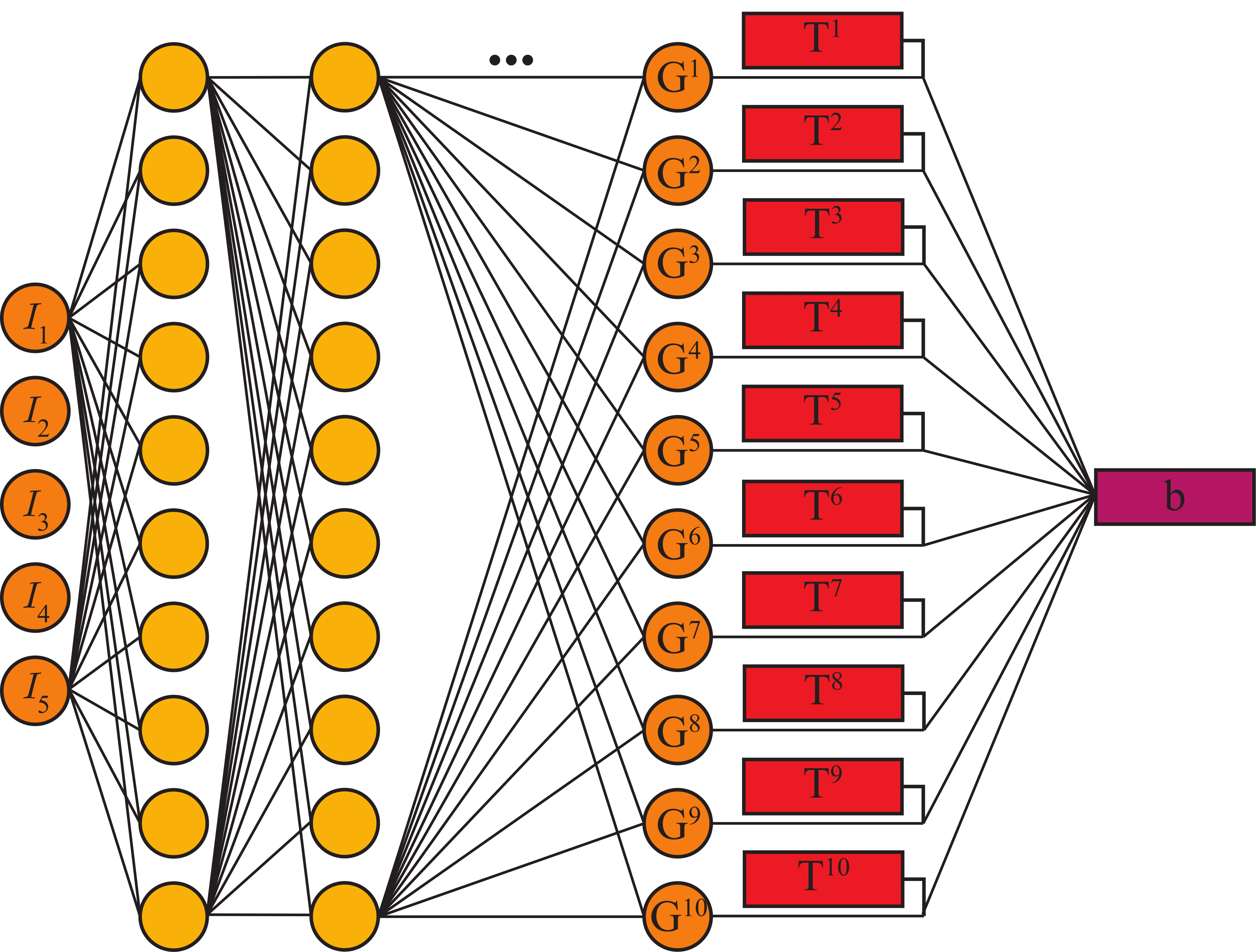}
    \caption{Invariant, fully-connected (some connections are omitted for clarity), neural network architecture proposed by Ling~\etal~\cite{ling2016reynolds}. The circles indicate scalar values and the rectangles represent $3 \times 3$ second-order tensors.}
    \label{fig:invarnn}
\end{figure}

We select a Bayesian neural network to map the baseline RANS flow to a high-fidelity R-S field due to the impressive performance of neural networks for high-dimensional supervised learning tasks~\cite{lecun2015deep}.
For the underlying neural network model, we choose the neural network proposed by Ling \etal~\cite{ling2016reynolds}, illustrated in Fig.~\ref{fig:invarnn}.
This neural network predicts the anisotropic tensor of the R-S using the symmetric and antisymmetric tensor components  of the velocity gradient tensor.
Through  use of tensor invariants, the neural network is able to achieve both Galilean invariance as well as invariance to coordinate transformations.
This makes such a model attractive for predictions of flows that deviate in geometry from the training data.
Here, we briefly review the fundamentals of this neural network for completeness of the presentation.

The theoretical foundation of this invariant neural network is the non-linear eddy viscosity model developed by Pope~\cite{pope1975more}.
In this model, the normalized anisotropic tensor of the R-S is expressed as a function, $\bm{b}(\bm{s},\bm{\omega})$, of the normalized mean rate-of-strain tensor $\bm{s}$ and rotation tensor $\bm{\omega}$:
\begin{gather}
    \left<u'_{i}u'_{j}\right>= \frac{2}{3}\delta_{ij}k + k  b_{ij}(\bm{s},\bm{\omega}),\label{eq:pope-rs}\\
    s_{ij} = \frac{1}{2}\frac{k}{\epsilon}\left(\frac{\partial\left<u_{i}\right>}{\partial x_{j}}+\frac{\partial \left<u_{j}\right>}{\partial x_{i}}\right), \quad
    \omega_{ij} = \frac{1}{2}\frac{k}{\epsilon}\left(\frac{\partial\left<u_{i}\right>}{\partial x_{j}} - \frac{\partial \left<u_{j}\right>}{\partial x_{i}}\right),
\end{gather}
where both $\bm{s}$ and $\bm{\omega}$ are scaled by the TKE and turbulent dissipation.
For clarity we will refer to the tensor, $\bm{a}=k\cdot\bm{b}(\bm{s},\bm{\omega})$, used when solving the RANS equations as the unnormalized anisotropic tensor.
Through application of the Cayley-Hamilton theorem, it can be shown that every second-order anisotropic tensor can be expressed in the following form:
\begin{gather}
    \bm{b}(\bm{s},\bm{\omega})=\sum_{\lambda=1}^{10}G^{\lambda}\left(\mathcal{I}_{1:5}\right)\bm{T}^{\lambda}, \label{eq:geneddypoly} \\ 
    \mathcal{I}_{i} = \left\{Tr(\bm{s}^{2}),\, Tr(\bm{\omega}^{2}),\, Tr(\bm{s}^{3}),\, Tr(\bm{\omega}^{2}\bm{s}),\, Tr(\bm{\omega}^{2}\bm{s}^{2})\right\}, \label{eq:geneddyinvar} \\
    \begin{aligned}
        \bm{T}^{1} & = \bm{s}, & \bm{T}^{2} & = \bm{s}\bm{\omega} - \bm{\omega}\bm{s}, & \bm{T}^{3} &= \bm{s}^{2} - \frac{1}{3}\bm{I}\text{Tr}\left(\bm{s}^{2}\right),\\
        \bm{T}^{4} &= \bm{\omega}^{2} - \frac{1}{3}\bm{I}\text{Tr}\left(\bm{\omega}^{2}\right), & \bm{T}^{5} &= \bm{\omega}\bm{s}^{2} - \bm{s}^{2}\bm{\omega}, & \bm{T}^{6} & = \bm{\omega}^{2}\bm{s} + \bm{s}\bm{\omega}^{2} - \frac{2}{3}\bm{I}\text{Tr}(\bm{s}\bm{\omega}^{2}),\\
        \bm{T}^{7} &= \bm{\omega}\bm{s}\bm{\omega}^{2} - \bm{\omega}^{2}\bm{s}\bm{\omega}, & \bm{T}^{8} &= \bm{s}\bm{\omega}\bm{s}^{2} - \bm{s}^{2}\bm{\omega}\bm{s}, & \bm{T}^{9} & = \bm{\omega}^{2}\bm{s}^{2} + \bm{s}^{2}\bm{\omega}^{2} - \frac{2}{3}\bm{I}\text{Tr}(\bm{s}^{2}\bm{\omega}^{2}), \\
        \bm{T}^{10} & = \bm{\omega}\bm{s}^{2}\bm{\omega}^{2} - \bm{\omega}^{2}\bm{s}^{2}\bm{\omega}, \label{eq:geneddytensor}
    \end{aligned}
\end{gather}
where $\bm{T}^{\lambda}$ is one of $10$ independent, symmetric tensor functions and $G^{\lambda}$ are the respective coefficients in the linear model which can be each expressed as  functions of the five invariants $\mathcal{I}_1,\cdots,\mathcal{I}_5$.
For complete details on the invariants, tensor functions and the derivation of the 
representation above, we refer the reader to~\cite{pope1975more}.

The neural network model proposed by Ling \etal~\cite{ling2016reynolds} models the anisotropic term by using the linear combination in Eq.~\eqref{eq:geneddypoly}.
As illustrated in Fig.~\ref{fig:invarnn}, rather than using the components of the symmetric and antisymmetric tensors ($\bm{s}$ and $\bm{\omega}$) directly, the invariants and tensor basis functions in Eqs.~\eqref{eq:geneddyinvar} and~\eqref{eq:geneddytensor} are used instead.
To enforce invariance to coordinate transformations, the neural network is used to learn the tensor basis coefficients $G^{\lambda}$ which are functions of the five invariants in Eq.~\eqref{eq:geneddyinvar}.
These predicted coefficients, $G^{\lambda}$, can then be used with the tensor basis functions, $\bm{T}^{\lambda}$, to produce the anisotropic tensor $\bm{b}$.
Thus while the model predicts the anisotropic tensor given the symmetric and antisymmetric tensors of the velocity gradient, the basis coefficients as functions of the five invariants is what is being learned.  
If a model uses inputs with specific invariant properties, the model has the same invariance properties as well~\cite{bishop2006machine}.
This allows the neural network to be (a) Galilean invariant due to the use of the rate-of-strain and rotation tensors which are functions of the velocity gradient; and (b) invariant to coordinate transformations through the use of the invariant inputs $\mathcal{I}_{i}$.
Additionally, since this eddy viscosity model is the most general formulation, this neural network model does not share any of the limitations of other simpler models that place restrictions on the form of the anisotropic term.
However, an intrinsic assumption of this model is that the mapping between the RANS and LES physical domains can be thoroughly expressed by the invariants $\mathcal{I}_{i}$.
This is clearly not guaranteed, however, the introduction of additional input features would potentially result in  loss of coordinate system invariance thus degrading model generalization.

\begin{remark}
This neural network formulation is trained on entirely local (point-wise) information.
The key advantage of a spatially local model is that it extends very easily to training flow data provided on non-uniform meshes which are essential in practical CFD simulations.
Approaches such as convolution neural networks require training data on a uniform mesh following an image-to-image like regression approach~\cite{zhu2018bayesian}.
However, similar to  turbulent eddy viscosity models, this approach implies that the R-S mean convection $D\left<u'_{i}u'_{j}\right>/Dt$ is governed entirely by local quantities (\textit{e.g.} $k,\,\epsilon,\, \partial \left<u_i\right>/\partial x_{j}$). This is a questionable assumption for flows that exhibit strong inhomogeneity~\cite{pope2001turbulent}.
A model that incorporates spatial correlations would likely be more descriptive, physically robust and potentially easier to train.
\end{remark}

\subsection{Bayesian Neural Network} \noindent
\label{subsec:svgd}
Traditionally neural networks are not designed to yield predictive statistics, however multiple recent works  explore Bayesian reformulations of neural networks.
Older techniques for obtaining Bayesian statistics include the placement of distributions over network weights and sampling with Monte Carlo methods to approximate statistical bounds~\cite{mackay1992bayesian, neal2012bayesian} as well as ensemble methods~\cite{richard1991neural,barber1998ensemble}.
More recently, methods involving stochastic variational inference have brought a new wave a Bayesian neural network techniques~\cite{blundell2015weight, kingma2015variational, gal2016dropout, liu2016stein}.
In this work, we choose to use Stein variational gradient decent (SVGD) recently proposed by Liu~\etal~\cite{liu2016stein, liu2017stein} that approximates a variational distribution through a set of particles.
SVGD is a non-parametric algorithm of similar form as  standard gradient decent.

We follow closely  the work of Zhu and Zabaras~\cite{zhu2018bayesian} in which SVGD is successfully applied to deep convolutional neural networks used for surrogate modeling.
For the invariant neural network architecture discussed previously, 
we will use the following representation:
\begin{equation}
    \bm{b} = \bm{f}(\left\{\bm{s}, \bm{\omega}\right\}, \mathbf{w}) = \bm{f}(\bm{x}, \mathbf{w}), \label{eq:nnFunc}
\end{equation}
where the input $\bm{x}=\{\bm{s},\bm{\omega}\}$ consists of the strain and rotation tensors $\bm{s}$ and $\bm{\omega}$ along with the neural networks parameters $\mathbf{w}$ which include  weights and biases.
For mathematical convenience, we will represent the anisotropic tensor with a one-dimensional vector  $\bm{b}\in\mathbb{R}^{9}$ for the remainder of this section. 
In the equation above, we have defined the neural network model as a function that has absorbed the calculation of the invariants, tensor basis functions and the linear combination detailed in Eqs.~\eqref{eq:geneddypoly}-\eqref{eq:geneddytensor}.
Thus we will refer to the function $\bm{f}$ as the invariant neural network model.
We wish to treat the neural network's $K$ learnable parameters as random variables.
Due to the potentially large number of weights in a fully-connected neural network, we assume that the weights have a probability density function of a fully-factorizable zero mean Gaussian and Gamma-distributed precision scalar $\alpha$:
\begin{equation}
    p(\mathbf{w}|\alpha)=\mathcal{N}(\mathbf{w}|0, \alpha^{-1}\bm{I}_{K}), \quad p(\alpha) = Gamma(\alpha|a_{0},b_{0}),
\end{equation}
where the rate  $a_{0}$ and  shape parameters $b_{0}$ are taken as $1.0$ and $0.025$,  respectively and $\bm{I}_{n}$ denotes the identity matrix in $\mathbb{R}^{n\times n}$.
The resulting prior  has the density of a narrow Student's $\mathcal{T}$-distribution centered at zero. This promotes sparsity~\cite{tipping2001sparse} and helps to prevent over-fitting.
With the use of a sufficient number of weights and the highly non-linear nature of the neural network, such a prior places little restriction on the  network's final functional form~\cite{neal2012bayesian}.
Additionally, output-wise noise is added onto the predicted output to represent inherent uncertainty within the model's formulation or uncertainty that cannot be reduced with more training data.
This results in an additional noise term to the likelihood function of the neural network.
We assume that the noise takes the form of a zero mean Gaussian with a learnable precision $\beta$ that is Gamma distributed:
\begin{gather}
    \bm{b} = \bm{f}(\bm{x}, \mathbf{w}) + \bm{\epsilon} \label{eq:nn-noise}, \\
    p(\bm{\epsilon}) = \mathcal{N}(\bm{\epsilon} | 0,\beta^{-1}\bm{I}_{9}), \quad p(\bm{b}) = \mathcal{N}(\bm{b}|\bm{f}(\bm{x}, \mathbf{w}),\bm{I}_{9}), \\
    p(\beta) = Gamma(\beta| a_{1}, b_{1}).
\end{gather}
Since both the LES and RANS simulations are being used for the same flows, we assume that the LES solution will be statistically stationary.
We also assume that the LES data has sufficiently converged by averaging over an adequate number of time steps.
The output-size noise is assumed to have a small variance and thus we assign in the prior for $\beta$ the shape and rate parameters to be $a_{1}=100$ and $b_{1}=2\cdot10^{-4}$, respectively. 
This weakly promotes large $\beta$ with an expected value of $5\times10^{5}$ and a variance on the order of $10^{-3}$, which is less than one percent of the scaled training data range.
For the sake of brevity, we will drop the notation of the conditional dependence on the hyper-parameters $a_{0}$, $b_{0}$, $a_{1}$ and $b_{1}$ implying that the posterior distribution will be conditionally dependent on these terms.
To optimize the parameters in the neural network, SVGD minimizes the Kullback-Leibler (KL) divergence between the true parameter posterior, $p(\mathbf{w},\beta|\mathcal{D})$, given the batch of $M$ i.i.d. training data $\mathcal{D}=\left\{\bm{b}_{i}\right\}_{i=1}^{M}$, with the variational distribution $q(\mathbf{w},\beta)$ that lies in some set of distributions $\mathcal{Q}$:
\begin{equation}
    q^{*}(\mathbf{w},\beta)=\min\limits_{q\in\mathcal{Q}}
    \left\{\text{KL}\left(q||p\right)\equiv\mathbb{E}_{q}(\log q(\mathbf{w},\beta)) - \mathbb{E}_{q}(\log \widetilde{p}(\mathbf{w},\beta| \mathcal{D})) + \mathcal{K}\right\},
\end{equation}
for which $\widetilde{p}(\mathbf{w},\beta|\mathcal{D})$ is the unnormalized posterior and $\mathcal{K}$ is the log normalization constant that is not required to be computed during optimization.
For the given neural network, we prescribe a Gaussian likelihood function and the priors discussed previously:
\begin{gather}
    \widetilde{p}(\mathbf{w},\beta| \mathcal{D}) = p(\mathcal{D}|\mathbf{w},\beta)p(\mathbf{w},\beta),\\
    \widetilde{p}(\mathbf{w},\beta| \mathcal{D}) = \prod^{M}_{i=1}\left[\mathcal{N}(\bm{b}_{i}|\bm{f}(\bm{x}_{i}, \mathbf{w}), \beta^{-1}\bm{I}_{9})\right]
    \mathcal{N}(\mathbf{w}|0, \alpha^{-1}\bm{I}_{K})\Gamma(\alpha| a_{0}, b_{0})\Gamma(\beta| a_{1}, b_{1}).\label{eq:likelihood}
\end{gather}
Rather than attempting to recover a parametric form of the variational distribution, SVGD describes $q(\mathbf{w}, \beta)$ by a particle approximation.
Namely, a set of $N$ deterministic neural networks each representing a particle $\left\{\bm{\theta}_{i}\right\}^{N}_{i=1}, \, \bm{\theta}_{i}=\left\{\mathbf{w}_{i}, \beta_{i}\right\}$, leading to an empirical measure $q_{N}(\mathbf{w}', \beta')=q_{N}(\bm{\theta}')=\frac{1}{N}\sum_{i=1}^{N}\delta(\bm{\theta}_{i}-\bm{\theta}')$.
Thus the objective is now for the empirical probability measure, $\mu_{N}$, to converge in distribution towards the true measure of the posterior $\nu$,
\begin{gather}
    \mu_{N}(d\bm{\theta})=\frac{1}{N}\sum_{i=1}^{N}\delta(\bm{\theta}_{i}-\bm{\theta})d\bm{\theta}=\frac{1}{N}\sum_{i=1}^{N}\bm{\theta}_{i}, \\
    \nu(d\bm{\theta})=p(\bm{\theta}|\mathcal{D})d\bm{\theta}.
\end{gather}
To minimize the KL divergence, we assume that $q(\mathbf{w}, \beta)$ is from a class of distributions that can be obtained through a set of smooth transforms.
A small perturbation function, resembling that of standard gradient decent, is used to update the particles:
\begin{equation}\label{eq:steinUpdate}
    \bm{\theta}^{t+1}_{i} = \bm{T}(\bm{\theta}^{t}_{i}) = \bm{\theta}_{i}^{t}+\eta^{t}\bm{\phi}(\bm{\theta}^{t}_{i}),
\end{equation}
where $\eta$ is the step size and $\bm{\phi}(\bm{\theta}_{i}^{t})$ is the direction of the update that lies in a function space $\mathcal{F}$ for the $t$-th iteration.
It is now a matter of finding the optimal direction to permute the particles which should be chosen such that the KL divergence is maximally reduced, namely,
\begin{equation}
    \bm{\phi}^{*} = \max_{\bm{\phi}\in\mathcal{F}}\left(-\frac{d}{d\eta}\mathcal{KL}( \bm{T}\mu_{N}||\nu)|_{\eta=0}\right),
\end{equation}
where $\bm{T}\mu$ denotes the updated empirical measure of the particles.
Liu \etal~\cite{liu2016stein} identify connections between the function $\bm{\phi}$ and Stein's method and show that:
\begin{equation}
    \frac{\partial}{\partial \eta}KL(\bm{T}\mu_{N}||\nu)|_{\eta = 0} = \mathbb{E}_{\mu}\left(\mathcal{T}_{p}\bm{\phi}\right), \quad \mathcal{T}_{p}\bm{\phi} = \left(\nabla \log p(\bm{\theta}| \mathcal{D})\right)\cdot\bm{\phi} + \nabla\cdot\bm{\phi},
\end{equation}
in which $\mathcal{T}_{p}$ is known as the Stein's operator.
Assuming that this function space $\mathcal{F}$ is a unit ball in a reproducing kernel Hilbert space $\mathcal{H}$ with positive kernel $k(\bm{\theta}, \bm{\theta}^{'})$, the optimal direction has the closed form:
\begin{equation}\label{eq:steinOpt}
    \bm{\phi}^{*}(\bm{\theta}) \propto \mathbb{E}_{\bm{\theta}^{\prime}\sim\mu}\left[\left(\nabla_{\bm{\theta}^{\prime}} \log p(\bm{\theta}^{\prime}| \mathcal{D})\right)k(\bm{\theta}, \bm{\theta}^{\prime}) + \nabla_{\bm{\theta}^{\prime}}k(\bm{\theta}, \bm{\theta}^{\prime})\right],
\end{equation}
where $p(\bm{\theta}^{\prime}| \mathcal{D})$ is given by Eq.~\eqref{eq:likelihood}.
In this work, we choose to use the standard radial basis function kernel for $k(\bm{\theta}, \bm{\theta}^{\prime})$.
This formulation results in a simple update procedure in which the optimal decent direction for all particles is calculated with Eq.~\eqref{eq:steinOpt} and then updated by Eq.~\eqref{eq:steinUpdate}.
Monte Carlo approximations can then be used to find the predictive mean:
\begin{align}
    \begin{split}
    &\mathbb{E}(\bm{b}^{*}|\bm{x}^{*},\mathcal{D})=\mathbb{E}_{p(\mathbf{w},\beta|\mathcal{D})}(\mathbb{E}(\bm{b}^{*}|\bm{x}^{*},\textbf{w},\beta))\\
    &\qquad=\mathbb{E}_{p(\mathbf{w}|\mathcal{D})}(\bm{f}(\bm{x}^{*},\mathbf{w}))\approx \frac{1}{N}\sum_{i=1}^{N}\bm{f}(\bm{x}^{*},\mathbf{w}_{i}),
    \label{eq:svgd-mean}
    \end{split}
\end{align}
where $\bm{x}^{*}$ and $\bm{b}^{*}$ are the test input and corresponding predictive model output, respectively.
The output noise is not present due to its density of a zero mean Gaussian.
The approximation of the predictive variance similarly follows:
\begin{align}
    \begin{split}
     \text{Cov}(\bm{b}^{*}|\bm{x}^{*},\mathcal{D})&=\mathbb{E}_{p(\mathbf{w},\beta|\mathcal{D})}(\text{Cov}(\bm{b}^{*}|\bm{x}^{*},\textbf{w},\beta)) + \text{Cov}_{p(\mathbf{w},\beta|\mathcal{D})}(\mathbb{E}(\bm{b}^{*}|\bm{x}^{*},\textbf{w},\beta))\\
     &=\mathbb{E}_{p(\beta|\mathcal{D})}(\beta^{-1}\bm{I}_{9}) + \text{Cov}_{p(\mathbf{w}|\mathcal{D})}(\bm{f}(\bm{x}^{*},\mathbf{w}))\\
     &\begin{aligned}
         &\approx\frac{1}{N}\sum_{i=1}^{N}\left((\beta_{i})^{-1}\bm{I}_{9}+\bm{f}(\bm{x}^{*},\mathbf{w}_{i})\bm{f}^{T}(\bm{x}^{*},\mathbf{w}_{i})\right)\\
         &\qquad\qquad-\mathbb{E}_{p(\mathbf{w}|\mathcal{D})}(\bm{f}(\bm{x}^{*},\mathbf{w}))\mathbb{E}^{T}_{p(\mathbf{w}|\mathcal{D})}(\bm{f}(\bm{x}^{*},\mathbf{w})), \label{eq:svgd-cov}
     \end{aligned}
     \end{split}
\end{align}
in which $\mathbb{E}_{p(\mathbf{w}|\mathcal{D})}(\bm{f}(\bm{x}^{*},\mathbf{w}))$ is calculated in Eq.~\eqref{eq:svgd-mean}.
Although the focus of this paper is to investigate the effect of the model form uncertainty on fluid quantities, the Bayesian neural network also allows for rigorous study of the epistemic uncertainty with Eqs.~(\ref{eq:svgd-mean}) and~(\ref{eq:svgd-cov}).
Thus one can study the effect of training data, model architecture, and other parameters on predictive confidence.
For complete details on SVGD, we direct the reader to the original work by Liu~\etal~\cite{liu2016stein, liu2017stein} along with the work of Zhu and Zabaras~\cite{zhu2018bayesian}.

\subsection{Uncertainty Quantification with SDD-RANS}
\label{subsec:uq}
\noindent
We now wish to propagate this uncertainty obtained for the anisotropic term to the fluid properties such as pressure or velocity.
We use a stochastic system approach for which the model parameters in a system of PDEs are considered as random variables.
This methodology has been used extensively in the past for model calibration, prediction and selection~\cite{beck1998updating, beck2002bayesian, cheung2009bayesian, cheung2011bayesian}.
Consider a dynamical system defined by the model output $h\left(\bm{\phi},\bm{u}(\bm{\phi})\right)$, where 
$\bm{u}(\bm{\phi})$ are state variables that evolve with the dynamical system and $\bm{\phi}$ represents a set of model parameters with probability density  $p(\bm{\phi})$. 
Traditionally the true form of the distribution $p(\bm{\phi})$ from which the model parameters are sampled from is largely not known.
However, under the assumption that samples can be drawn from the parameter distribution, the expected response as well as the respective variance can be approximated by vanilla Monte Carlo simulation (MCS) with $P$ samples of the random model parameters:
\begin{gather}
    \mathbb{E}_{p(\bm{\phi})}(h) \approx \frac{1}{P}\sum_{i=1}^{P}h\left(\bm{\phi}_{i},\bm{u}(\bm{\phi}_{i})\right), \quad \bm{\phi}_{i}\sim p(\bm{\phi}), \\
    \text{Var}_{p(\bm{\phi})}(h) \approx \frac{1}{P}\sum_{i=1}^{P}\left[h\left(\bm{\phi}_{i},\bm{u}(\bm{\phi}_{i})\right) - \mathbb{E}_{p(\bm{\phi})}(h)\right]^{2}.
\end{gather}

To extend this to the problem of interest and motivate SDD-RANS, let us consider the model output $h$ as the flow field predicted by the RANS equations and the state variables $\bm{u}(\bm{\phi})$ to be the fluid's velocity, pressure and all other derived properties.
As previously discussed, we will be taking an explicit representation of the tuned R-S in which a modified R-S field is predicted and held constant while the other state variables are propagated forward.
Thus we are able to view a predicted R-S field as a random model parameter, namely, $p(\bm{\phi}) = p(\bm{b}^{*}|\bm{x}^{*},\mathcal{D})$ where the predictive density of $\bm{b}^{*}$ is given by:
\begin{equation}
    p(\bm{b}^{*}|\bm{x}^{*},\mathcal{D}) = \int p(\bm{b}^{*}|\bm{x}^{*},\textbf{w},\beta)p(\textbf{w},\beta|\mathcal{D})d\textbf{w}d\beta.
\end{equation}
Rather than sampling the anisotropic term directly from the predictive distribution, recall the following representation of the likelihood in Eq.~\eqref{eq:nn-noise}:
\begin{equation*}
    \bm{b}^{*} = \bm{f}(\left\{\bm{s}^{*}, \bm{\omega}^{*}\right\}, \mathbf{w}) + \bm{\epsilon}.
\end{equation*}
To sample the predictive distribution, one can first sample the posterior $p(\textbf{w},\beta|\mathcal{D})$ and then execute a forward prediction of the neural network as well as sample the additive output noise yielding the predicted $\bm{b}^{*}$.
We can modify the MCS such that we sample the weights of the Bayesian neural network as well as the variance of the additive output-wise noise:
\begin{gather}
    \mathbb{E}_{p(\mathbf{w},\beta|\mathcal{D})}(h) \approx \frac{1}{N}\sum_{i=1}^{N}h\left(\bm{b}^{*}_{i},\bm{u}(\bm{b}^{*}_{i})\right), \\
    \text{Var}_{p(\mathbf{w},\beta|\mathcal{D})}(h) \approx \frac{1}{N}\sum_{i=1}^{N}\left[h\left(\bm{b}^{*}_{i},\bm{u}(\bm{b}^{*}_{i})\right) - \mathbb{E}_{p(\mathbf{w},\beta|\mathcal{D})}(h)\right]^{2},\\
    \bm{b}^{*}_{i} = \bm{f}(\left\{\bm{s}^{*}, \bm{\omega}^{*}\right\} , \mathbf{w}_{i}) + \bm{\epsilon}_{i}, \quad \bm{\epsilon}_{i} \sim \mathcal{N}(\bm{\epsilon}_{i} | 0,\beta_{i}^{-1}\bm{I}_{9}), \\
    \left\{\textbf{w}_{i}, \beta_{i}\right\} \sim p(\mathbf{w}_{i}, \beta_{i}|\mathcal{D}).
\end{gather}
The SVGD algorithm  provides samples of the  posterior $p(\mathbf{w}_{i}, \beta_{i}|\mathcal{D})$.
Namely, given that SVGD uses a particle representation, each sample is a particle (or invariant neural network) used during training.
In practice, the output-wise noise, $\bm{\epsilon}$, has minimal influence on the predicted values due to the previously made assumptions.
Hence, we only take a mean point estimate of the likelihood.
In principle, the neural network's inputs are spatially independent between mesh nodes allowing for each node point in the fluid domain to have independent weight samples resulting in a stochastic field.
However, the use of the divergence of the R-S in the RANS equations suggests that the spacial smoothness of the predicted field is of significant importance.
Thus, we use a single neural network, $\bm{f}(\left\{\bm{s}^{*}, \bm{\omega}^{*}\right\} , \mathbf{w}_{i})$, to predict the R-S for the entire flow domain.
As a result, in the context of the fluid domain, we are in fact sampling a functional representation of the R-S that is dependent on the velocity gradients.
This combination of using a Bayesian data-driven model with a stochastic model parameter is why we have named this process stochastic data-driven RANS (SDD-RANS).
With SDD-RANS, we have opened up the ability to obtain sample statistics for all flow quantities through traditional MCS.
This allows for the quantification of uncertainty regarding our data-driven model beyond the R-S itself.

The use of the explicit representation of the R-S and the noisy nature of the neural network's predictions raise  concerns regarding the convergence of the SDD-RANS model.
In practice, at higher Reynolds numbers the simulation may fail to converge in some areas of the domain.
Due to the nature of SDD-RANS, the statistical averages obtained through MCS accurately reflect the true state of the quantities of interest.
The discrepancy from the true solution is reflected by the computed variance or uncertainty estimates.
However, while computing each sample, one must still monitor the residuals of the model to ensure the initial transient state has ended.
In practice, we run the forward simulation for the same number of iterations as the baseline simulation.

\subsection{Framework Implementation} \noindent
We use this Bayesian framework in the system of RANS equations by setting the R-S term as the stochastic parameter that is sampled from the predictive distribution obtained through the Bayesian neural network.
We summarize the offline training process:
\begin{itemize}
    \item The training data consist of both baseline RANS and high-fidelity data for a set of different flows that attempt to capture different flow physics.
    \item The underlying machine learning model is an invariant neural network that uses local fluid quantities to predict the anisotropic term of the R-S.
    \item We extend this invariant model to the Bayesian paradigm by using SVGD in which a set of neural networks approximates the posterior $p(\mathbf{w},\beta|\mathcal{D})$ by a particle representation.
    \item The parameters in each particle (or neural network) are optimized by minimizing the KL divergence between a particle variational approximation and the posterior of the parameters.
    \item An iterative algorithm, resembling the form of standard gradient decent, updates the parameters of each particle until convergence.
\end{itemize}
With the Bayesian neural network trained, one can make predictions for new flows:
\begin{itemize}
    \item For the flow of interest, a baseline RANS solution is obtained and the corresponding invariants and tensor functions at each mesh point are calculated.
    \item Each neural network used during training with SVGD is used to predict a corresponding high-fidelity R-S field.
    \item For each predicted field, the R-S is then constrained to the predicted values and a forward execution of the constrained system updates the remaining state variables.
    \item An equivalent number of state variable samples are then obtained for which probabilistic bounds can be calculated.
\end{itemize}

\section{Numerical Implementation and Training}
\label{sec:Training}

\subsection{CFD Methods}
\noindent
For obtaining the training and test flows, the open source CFD platform OpenFOAM (Open source Field Operation And Manipulation)~\cite{weller1998tensorial, jasak2007openfoam} is used. 
OpenFOAM is a widely accepted CFD package that contains a vast number of solvers for incompressible, compressible and multi-phase flows along with pre- and post-processing utilities.
For the baseline RANS simulations the steady-state, incompressible solver \textit{simpleFoam} was used which employs the semi-implicit method for pressure linked equations (SIMPLE) algorithm~\cite{patankar1983calculation} to solve both the momentum and pressure equations.
The high-fidelity LES simulations used the \textit{pimpleFoam} transient solver that combines both the PISO (Pressure Implicit with Split Operator)~\cite{issa1986solution} and SIMPLE algorithms to solve the pressure and momentum equations.
The Smagorinsky subgrid-scale model~\cite{smagorinsky1963general} with Van-Driest style damping was used for all LES flows.
Both the baseline RANS and LES domains are discretized by second-order methods.
Each training and testing flow is outlined in Tables~\ref{tab:cfdflows} and~\ref{tab:cfd-test-flows}, and all meshes are non-uniform such that the mesh density increases around the feature of interest.
All simulations were run with a CFL number below $0.3$ for numerical accuracy.

\newgeometry{margin=1cm} 
\thispagestyle{empty}
\begin{landscape}
\begin{table}[t]
\centering
\caption{Mesh and CFD parameters for each training flow which includes the respective reference, mesh sizes for both RANS and LES simulations, the domain size, characteristic length $L_{c}$, bulk Reynolds number $Re_{b}$ and kinematic viscosity $\nu$. Streamwise is in the $x-$, wall normal in the $y-$ and spanwise in the $z-$direction.}
\label{tab:cfdflows}
\begin{tabular}{p{3cm}ccccc}
\hline
Case                         & Converge Diverge & Square Cylinder & Periodic Hills & Square Duct & Tandem Cylinders \\ \hline
Reference & Schiavo~\etal~\cite{schiavo2015large, langley2018converge} & Bosch~\etal~\cite{bosch1998simulation} &  Temmerman~\etal~\cite{temmerman2003investigation, langley2018periodic} & Pinelli~\etal~\cite{pinelli2010reynolds} &  Gopalan~\etal~\cite{gopalan2015numerical} \\
Mesh RANS                    & $140 \times 50 \times 50$                                  & $100 \times 60 \times 20$                      & $100 \times 50 \times 50$                      & $7.5 \pi  \times 60 \times 60$                 & $80 \times 60 \times 50$        \\
Mesh LES                     & $280 \times 100 \times 150$                                & $280 \times 120 \times 40$                     & $500 \times 150 \times 250$                    & $300 \pi  \times 150 \times 150$               & $470 \times 180 \times 120$     \\
Domain Size                  & $12.56H \times 2H \times 3H$                               & $20D \times 14D \times 4D$                      
& $9H \times 3.306H \times 4.5H$                 & $4 \pi H \times 2H \times 2H$                  & $30D \times 20D \times 3D$      \\
$L_{c}$        & Half Channel Height                            & Cylinder Diameter                   & Hill Height                        & Half Channel Width              & Cylinder Diameter      \\
$Re_{b}$       & $5000$                                           & $5000$                                & $6210$                               & $6680$                            & $5000$                   \\
$\nu$          & $2.00$e$-4$ & $2.00$e$-4$  & $6.07$e$-4$ & $2.00$e$-4$ & $7.40$e$-4$\\
\end{tabular}
\end{table}

\begin{table}[t]
\centering
\caption{Mesh and CFD parameters for each test flow  which includes the respective reference, mesh sizes for both RANS and LES simulations, the domain size, characteristic length $L_{c}$, bulk Reynolds number $Re_{b}$ and kinematic viscosity $\nu$. Streamwise is in the $x-$, wall normal in the $y-$ and spanwise in the $z-$direction.}
\label{tab:cfd-test-flows}
\begin{tabular}{lccc}
\hline
Case                         & Backward Step & Wall Mounted Cube \\ \hline
Reference &  Gresho~\etal~\cite{gresho1993steady}  & Yakhot~\etal~\cite{yakhot2006turbulent} \\
Mesh RANS                    & $220 \times 60 \times 20$                        & $100 \times 40 \times 80$                   \\
Mesh LES                     & $390 \times 100 \times 40$                       & $200 \times 100 \times 150$                 \\
Domain Size                  & $27H \times 2H \times H$                         & $14H \times 3H \times 7H$                   \\ 
$L_{c}$         & Step Height                          & Cube Height                     \\
$Re_{b}$       & $500, 2500, 5000$                      & $500, 2500, 5000$                 \\
$\nu$           & $2.00$e$-3$, $4.00$e$-4$, $2.00$e$-4$ & $2.00$e$-3$, $4.00$e$-4$, $2.00$e$-4$ \\
\end{tabular}
\end{table}

\end{landscape}
\restoregeometry

\subsection{Machine Learning Implementation}\label{sec:TrainingML}
\noindent
To train the neural network, the Python machine learning library PyTorch~\cite{paszke2017automatic} was used. 
The software and data used in this work are available at \href{https://github.com/cics-nd/rans-uncertainty}{https://github.com/cics-nd/rans-uncertainty}.
The details of the network architecture used are given in Table~\ref{tab:nn}.
The Leaky Rectifier function was used as the activation function as opposed to the standard Rectifier function to prevent too many nodes from becoming zero during training.
Additionally the number of nodes in the hidden layers is tapered at the end of the network to prevent weights from being too small, which improved training performance.

\begin{table}[H]
\centering
\caption{Neural network architecture and training details.}
\label{tab:nn}
\begin{tabular}{ll}
Architecture     & $5\rightarrow200\rightarrow200\rightarrow200\rightarrow40\rightarrow20\rightarrow10$             \\
Activation      & Leaky ReLu                                \\
Optimizer       & ADAM~\cite{kingma2014adam}                 \\
Weight Decay    & $0.01$  \\
Learning Rate   & $5$e$-6$, with learning rate decay on plateau \\
Epochs          & $100$                                        \\
Training Data   & $10000$                                    \\
Mini-batch size & $20$                                       \\
SVGD Particles & $20$                          
\end{tabular}
\end{table}

\begin{figure}[H]
    \centering
    \includegraphics[width=1.0\textwidth]{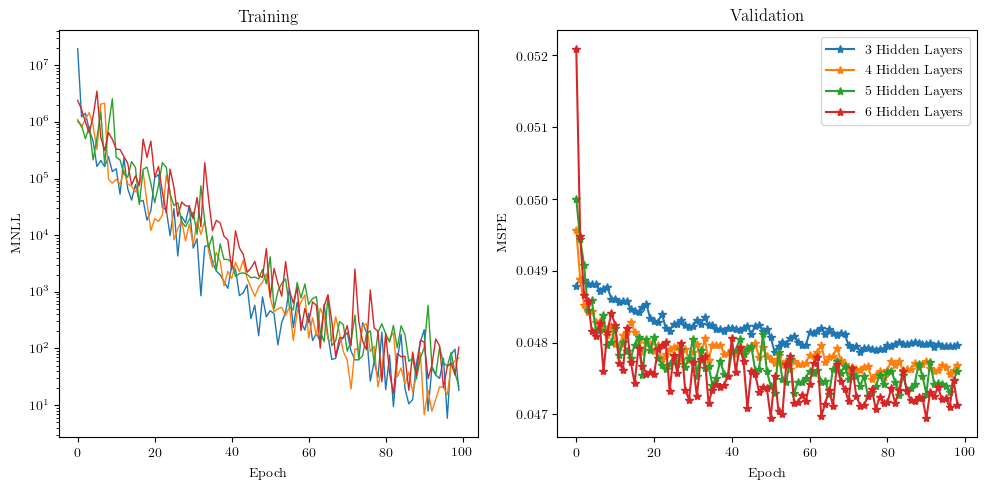}
    \caption{(Left) The negative log likelihood (MNLL) and (Right) the mean squared prediction error (MSPE) of the validation data for several neural network architectures.}
    \label{fig:hiddenLayersMSE}
\end{figure}

The network architecture was determined by training an ensemble of neural networks with different number of hidden layers.
Other network parameters, such as the taper of the last several layers and learning rate specified in Table~\ref{tab:nn}, were identical between each of the tested architectures.
The networks are compared in Fig.~\ref{fig:hiddenLayersMSE} with the mean negative log likelihood (MNLL) defined by:
\begin{gather}
    \quad {MNLL}=-\frac{1}{T}\sum_{i=1}^{T}\frac{1}{N}\sum_{j=1}^{N}\log{p(\hat{\bm{b}}_{i}|\bm{f}(\hat{\bm{x}}_{i}, \mathbf{w}_{j}), \beta_{j})},
\end{gather}
where $T$, $\hat{\bm{x}}$, $\hat{\bm{b}}$ are the number of validation/test data points, the target inputs and target outputs, respectively.
We observe little distinguishable difference indicating that training between each architecture is relatively the same.
Additionally, the mean squared prediction error of the unnormalized anisotropic tensor $\bm{a}^{*}$ is plotted for a validation set of $1000$ random data points from each of the training flows (i.e. $5000$ total data points).
The mean squared prediction error (MSPE) is defined as:
\begin{equation}
    {MSPE}=\frac{1}{T}\sum^{T}_{i=1}\norm{\mathbb{E}(\bm{a}^{*}_{i}|\hat{\bm{x}}_{i},\mathcal{D})-\hat{\bm{a}}_{i}}^{2}_{2}\approx\frac{1}{T}\sum^{T}_{i=1}\norm{\frac{k}{N}\sum_{j=1}^{N}\bm{f}(\hat{\bm{x}}_{i},\mathbf{w}_{j})-\hat{\bm{a}}_{i}}^{2}_{2},
\end{equation}
where $k$ is the baseline RANS TKE and $\hat{\bm{a}}$ are the target unnormalized anisotropic tensors.
For the MSPE, there is a notable difference between the converged accuracy of each network.
The networks with above $6$ hidden layers exhibited significant over-fitting of the validation data and are not plotted.
One can observe the onset of over-fitting by the noisy MSPE of the $6$ hidden layer neural network.
The network architecture with $5$ hidden layers was selected to ensure that over-fitting does not take place.

\begin{figure}[H]
    \centering
    \includegraphics[width=1.0\textwidth]{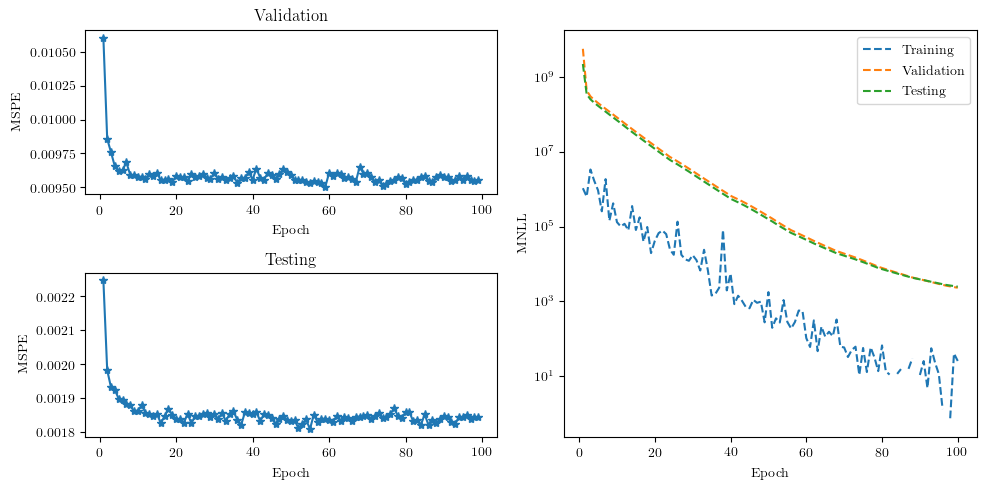}
    \caption{(Left) The mean squared prediction error (MSPE) of both the validation and test datasets. (Right) The mean negative log likelihood (MNLL) of the training, validation and test datasets.}
    \label{fig:testMSE}
\end{figure}

Compared to other potential network models, we found that the model selected originally by Ling~\etal~\cite{ling2016reynolds} proved to be exceptionally difficult to train.
This is reflected in the original work by the extremely low learning rate used of $2.5 \times 10^{-6}$.
During training we also found only very low learning rate could be used for the training process to be stable.
To increase training performance and efficiency, we also used the following techniques:
\begin{itemize}
    \item Depending on the size of the fluid domains, the number of training points for a single flow can be large.
    Thus to increase training efficiency, rather than using every single mesh point, a subset of training points is selected.
    In this work, we use only $10^4$ total training points that are evenly distributed among all training flows (i.e. $2 \times 10^3$ points for each of the five test flows).
    Every $10$ epochs these points are then re-sampled at random.
    This prevents the potential issue of exceeding the available memory on the provided GPU.
    \item Training points are shuffled randomly and mini-batched every epoch such that data from multiple flows can reside in a single mini-batch.
    This helps prevent the model from over-fitting to a specific flow and improves the quality of predictions.
    \item The invariant inputs to the neural network tended to vary strongly in magnitude including very large values near fixed boundaries.
    Thus the invariants are re-scaled by a sigmoidal operation that helps to normalize outliers to the range of $+1$ to $-1$~\cite{li2000fuzzy}. 
    In addition, the tensor basis functions were normalized by the $L_2$ norm of the matrix:
    \begin{equation}
        \hat{\mathcal{I}}_{i} = \frac{1-e^{-\mathcal{I}_{i}}}{1+e^{-\mathcal{I}_{i}}}, \qquad \hat{\bm{T}}^{\lambda} = \frac{\bm{T}^{\lambda}}{\norm{\bm{T}^{\lambda}}_{2}}.
    \end{equation}
\end{itemize}
\noindent
To quantify the training quality, the MSPE is calculated for both the validation set along with a test set of $1000$ randomly selected points from each test flow in Table~\ref{tab:cfd-test-flows}.
Additionally, we also plot the MNLL for the training, validation and testing data sets.
The results are illustrated in Fig.~\ref{fig:testMSE}.
We note that for both the validation and test datasets the model quickly converges and exhibits minimal over-fitting.
The training process on a single NVIDIA P100 GPU took approximately $3.0$ wall-clock hours.

\begin{figure}[H]
    \centering
    \includegraphics[width=1.0\textwidth]{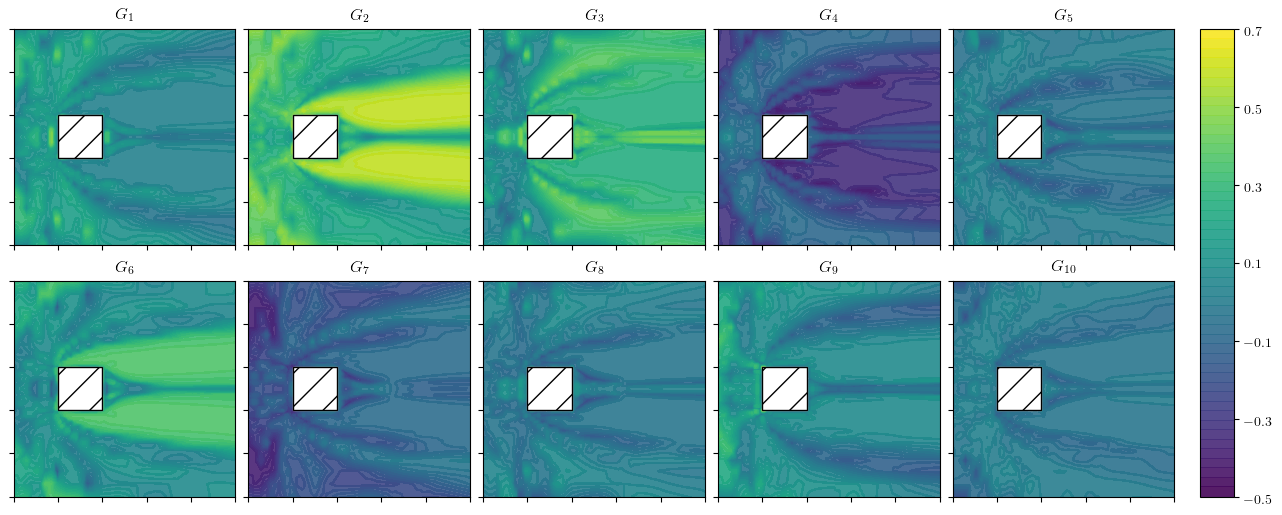}
    \caption{The learned mixing coefficients of the training neural network for flow around a square cylinder~\cite{bosch1998simulation} at bulk Reynolds number $5000$.
    No domain reflections were used to artificially impose symmetry.}
    \label{fig:squareGCoeff}
\end{figure}

\begin{figure}[H]
    \centering
    \includegraphics[width=1.0\textwidth]{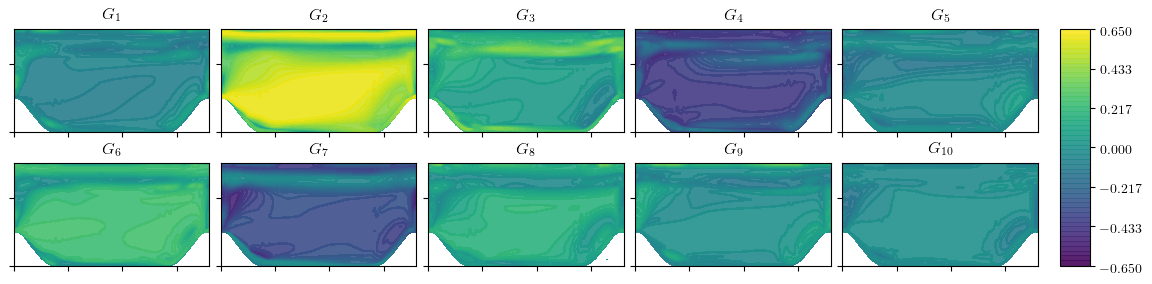}
    \caption{The learned mixing coefficients of the training neural network for flow over periodic hills~\cite{temmerman2003investigation, langley2018periodic} at bulk Reynolds number $6210$.}
    \label{fig:periodHillGCoeff}
\end{figure}

To verify that the trained model has learned a physical interpretation of the training data, we plot the contours of the mixing coefficients $G^{\lambda}$ predicted by the neural network for the square cylinder and periodic hills training flows in Figs.~\ref{fig:squareGCoeff} and~\ref{fig:periodHillGCoeff}, respectively.
While there appears to be some minor over-fitting in front of the   square cylinder in Fig.~\ref{fig:squareGCoeff},
for both flows it is clear that the model has indeed identified physical regions of the flow as well as maintained symmetries.

\subsection{Constrained R-S Simulation}
\noindent
To integrate the sampled R-S field into OpenFOAM, a small modification is made to the \textit{simpleFoam} solver such that the R-S is now a constant field in the momentum RANS equation.
Since the R-S field is held constant, the calculation of the TKE and turbulent dissipation is no longer needed.
An important issue is the handling of boundary conditions.
This includes the treatment of domain boundaries as well as areas in which wall functions may be used.
We address these issues using two different methods.
First, the use of the baseline RANS TKE as a scaling factor of the anisotropic term shown in Eq.~\eqref{eq:pope-rs} allows for many turbulent boundary conditions to be satisfied.
Second, to address areas in which wall functions may be used, we take inspiration from hybrid LES/RANS methods and introduce a blending function proposed by Xiao~\etal~\cite{xiao2004blending}:
\begin{equation}
    \bm{b}^{*} = \Gamma \bm{b}_{dd} + (1-\Gamma)\bm{b}_{rans}, \quad \Gamma=\tanh{\left(d/\alpha_{1}\lambda\right)}^{2}, 
\end{equation}
where $\bm{b}_{dd}$ is the data-driven prediction of the anisotropic tensor, $\bm{b}_{rans}$ is the baseline RANS anisotropic tensor, $\lambda^{2} = k/\epsilon$, $d$ is the distance from the wall and $\alpha_{1}$ is a tunable parameter.
This function allows a smooth transition between the use of the baseline R-S near the wall and the data-driven prediction in the bulk flow.
In the original work, it is suggested that the selection of $\alpha_{1}$ be a value that achieves $\Gamma = 0.5$ somewhere is the log region.
We found the value of $0.05$ worked well for our test cases.

\section{Numerical Results}
\label{sec:NumericalResults}
\begin{figure}[h]
    \centering
    \begin{subfigure}[b]{0.48\textwidth}
        \includegraphics[width=\textwidth]{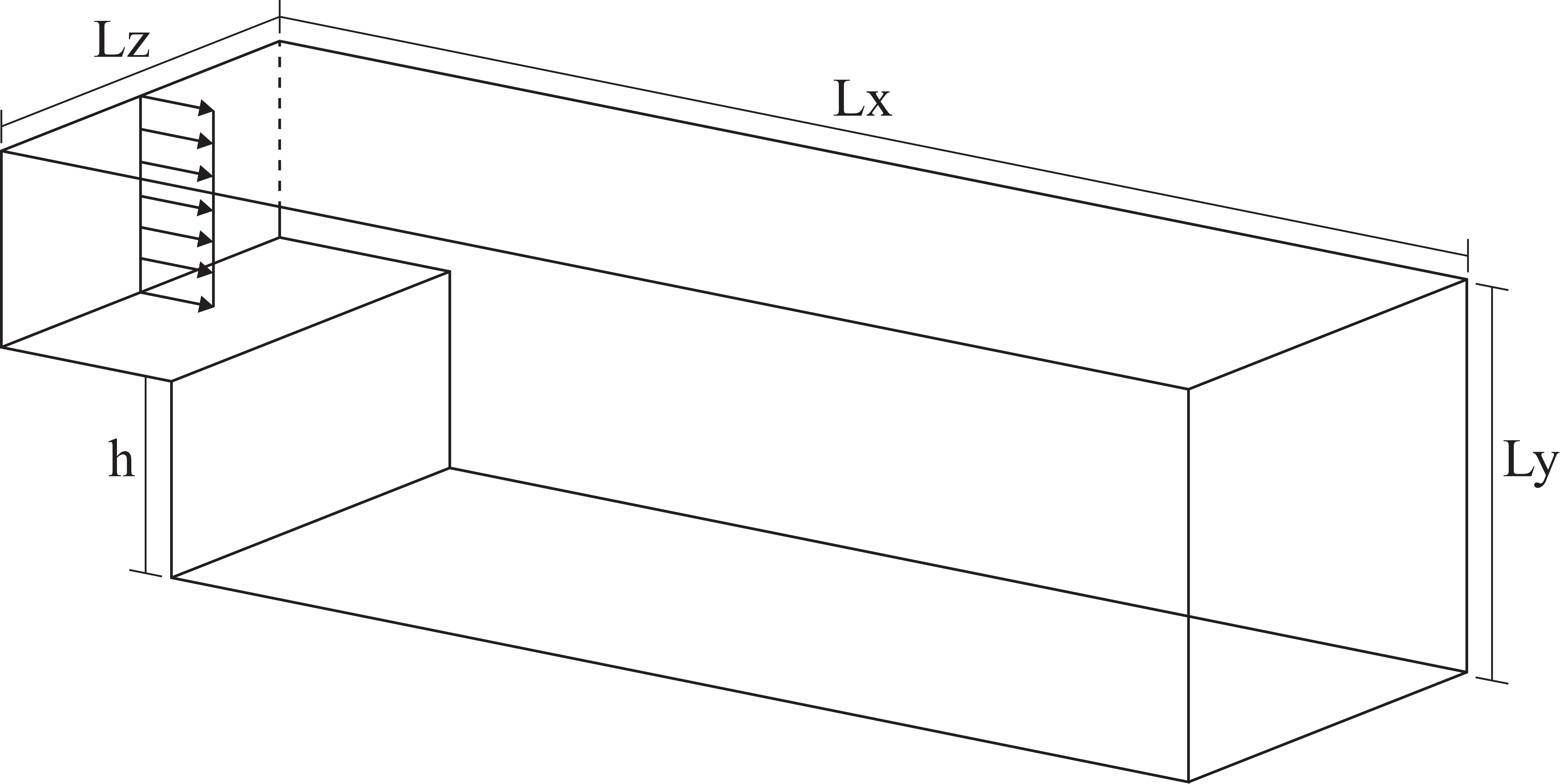}
        \caption{}
        \label{fig:backstepSchematic}
    \end{subfigure}
    ~
    \begin{subfigure}[b]{0.48\textwidth}
        \includegraphics[width=\textwidth]{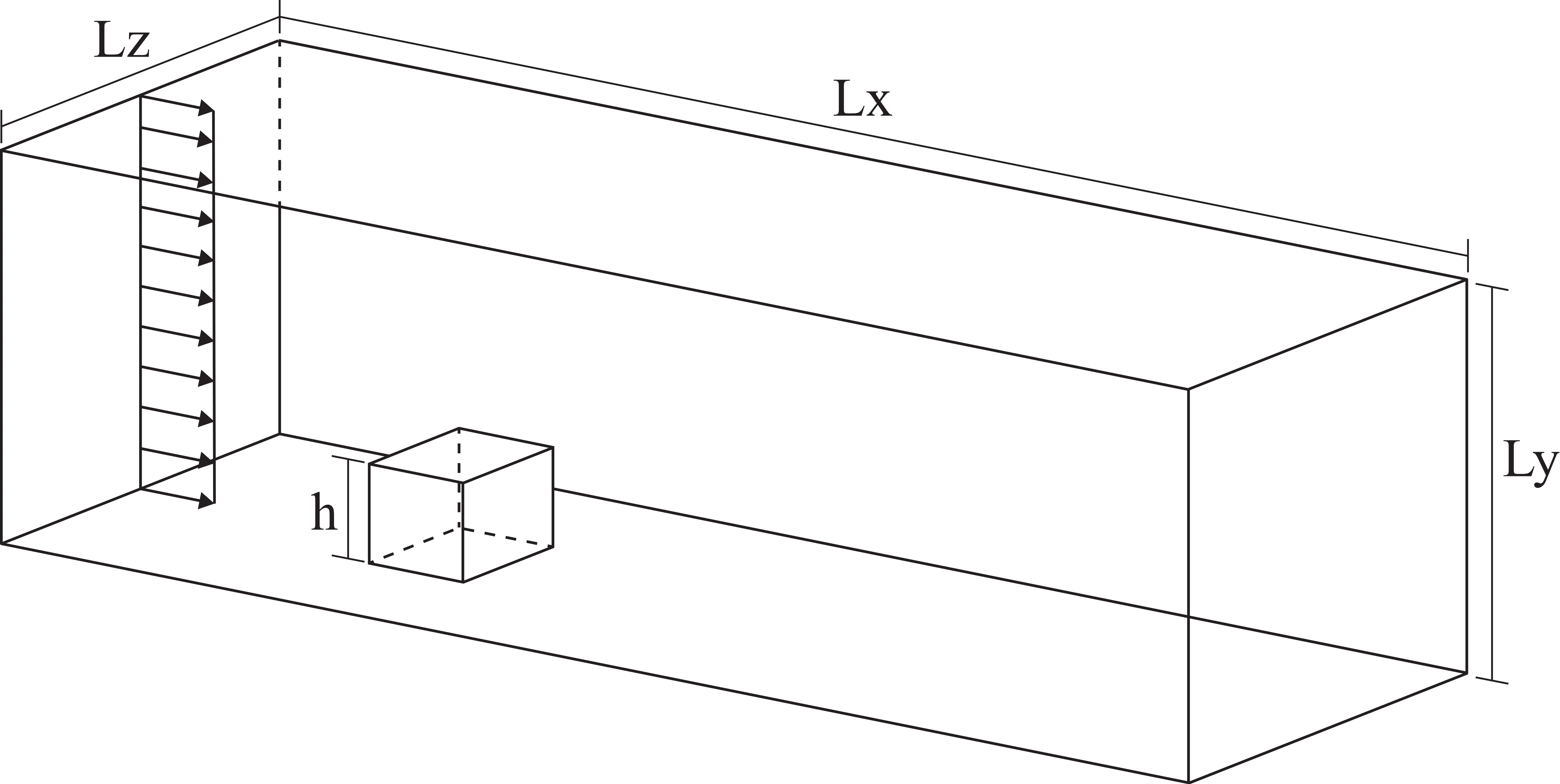}
        \caption{}
        \label{fig:wallCubeSchematic}
    \end{subfigure}
    \caption{(a) Flow geometry for the backwards step test flow with height $h$. (b) Flow geometry for the wall mounted cube test flow with height $h$.}
    \label{fig:testFlowSchematic}
\end{figure}

\noindent
The use of data-driven models for test simulations whose domain is similar or identical to the training data is a frequent occurrence in the literature but does not correctly assess a data-driven model's performance.
Since our selected neural network has already been shown to work adequately for similar flows in~\cite{ling2016reynolds}, our test cases are selected to deviate significantly from the training flows in both flow geometry and Reynolds number.
We have selected the two test flows detailed in Table~\ref{tab:cfd-test-flows}: flow over a backwards step and flow around a wall mounted cube both at three different Reynolds numbers.
The geometry for each flow can be seen in Fig.~\ref{fig:testFlowSchematic}.
For both test cases,  we will study the accuracy and respective uncertainty of both the model's R-S predictions as well as the predicted fluid quantities of interest.
Ultimately, we wish to assess the predictive performance of the data-driven model for these geometrically different flows and use the proposed stochastic data-driven RANS algorithm to calculate probabilistic bounds on flow state variables by 
conducting uncertainty quantification on the data-driven turbulence model.

\subsection{Backwards Step}
\label{subsec:backwardsStep}
\begin{figure}[H]
    \centering
    \includegraphics[width=0.95\textwidth]{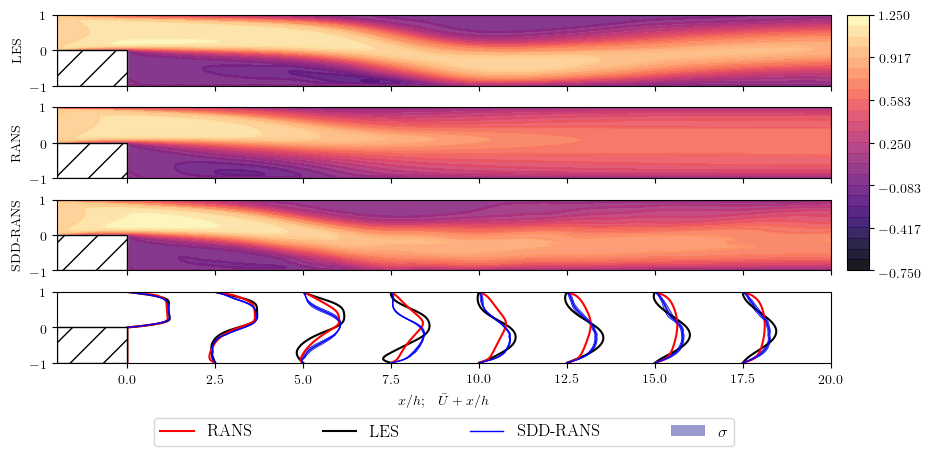}
    \caption{Normalized stream-wise mean velocity contours for Reynolds number $500$. The top is the LES solution, below is the baseline RANS prediction followed by the data-driven mean field. Lastly is the stream-wise mean velocity profiles for all simulations  shown along with the predictive error bars of the SDD-RANS prediction.}
    \label{fig:backstepUX500}
\end{figure}

\begin{figure}[H]
    \centering
    \includegraphics[width=0.95\textwidth]{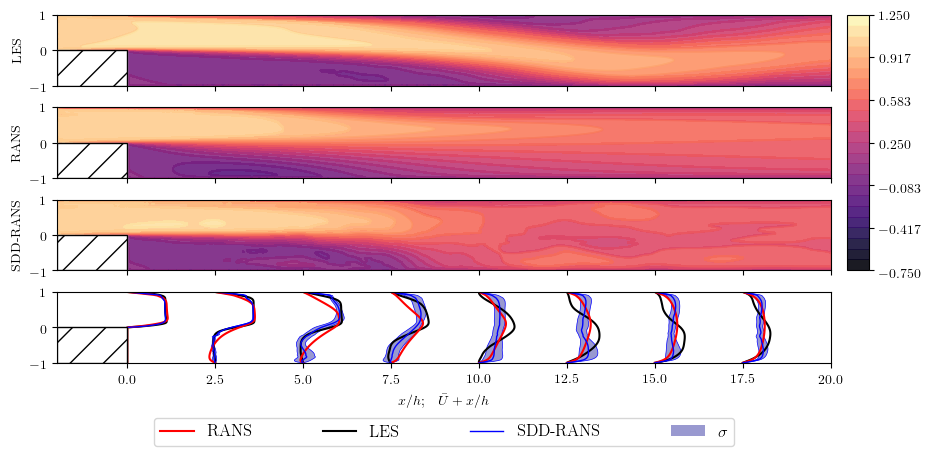}
    \caption{Normalized stream-wise mean velocity contours for Reynolds number $2500$. The top is the LES solution, below is the baseline RANS prediction followed by the data-driven mean field. Lastly is the stream-wise mean velocity profiles for all simulations  shown along with the predictive error bars of the SDD-RANS prediction.}
    \label{fig:backstepUX2500}
\end{figure}

\begin{figure}[H]
    \centering
    \includegraphics[width=0.95\textwidth]{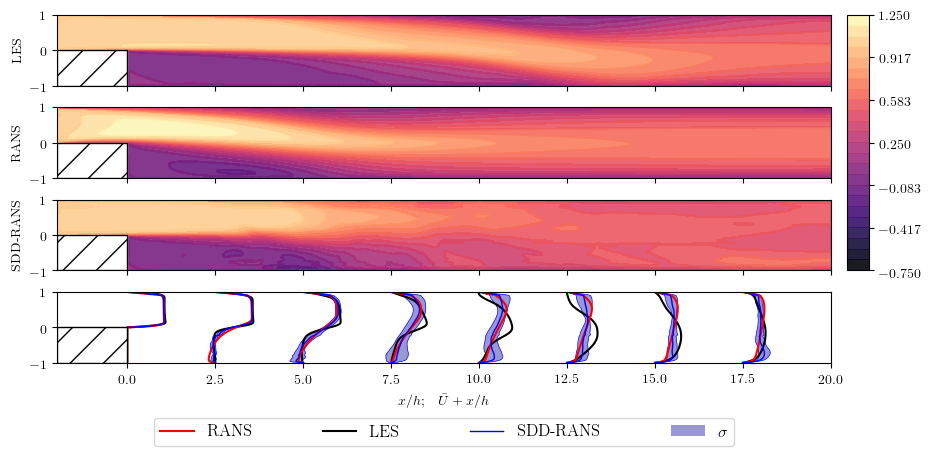}
    \caption{Normalized stream-wise mean velocity contours for Reynolds number $5000$. The top is the LES solution, below is the baseline RANS prediction followed by the data-driven mean field. Lastly is the stream-wise mean velocity profiles for all simulations  shown along with the predictive error bars of the SDD-RANS prediction.}
    \label{fig:backstepUX5000}
\end{figure}

\noindent
In the  first test case, we select a backwards facing step at three different Reynolds numbers.
As illustrated in Fig.~\ref{fig:backstepSchematic}, this flow features a constant velocity inlet channel followed by a backwards facing step of height $h$.
For this flow, we select the inlet channel to be the same height as the step.
In contrast to most backwards step simulations, no-slip walls are on both the top and bottom faces.
The $z$ direction is periodic.
On the $x-y$ plane, we place the origin at the corner of the step.
The flow features of interest are the recirculating regions that appear not only directly after the step but also on the upper channel wall down stream which is seen in the LES simulations in Figs.~\ref{fig:backstepUX500}-\ref{fig:backstepUX5000}.

\begin{figure}[H]
    \centering
    \includegraphics[width=0.9\textwidth]{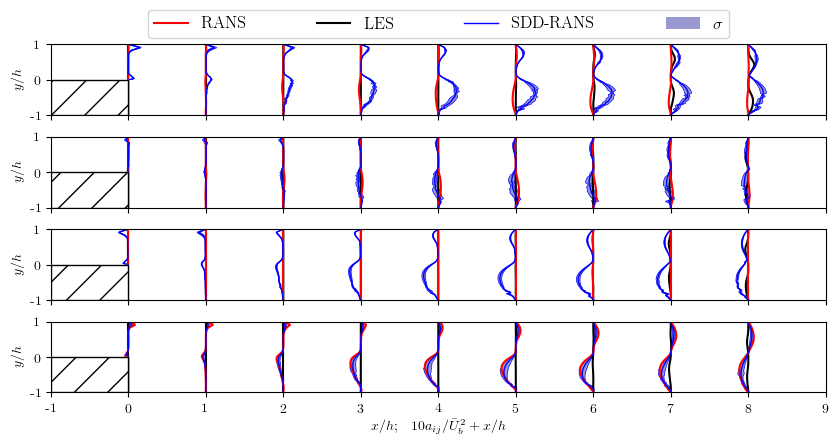}
    \caption{The anisotropic term predictions for the backwards step test flow for Reynolds number $500$. Top to bottom: $a_{11}$, $a_{22}$, $a_{33}$ and $a_{12}$ ($a_{13}$ and $a_{23}$ are omitted due to all fields being zero).}
    \label{fig:backstepDeviatoric500}
\end{figure}

\begin{figure}[H]
    \centering
    \includegraphics[width=0.9\textwidth]{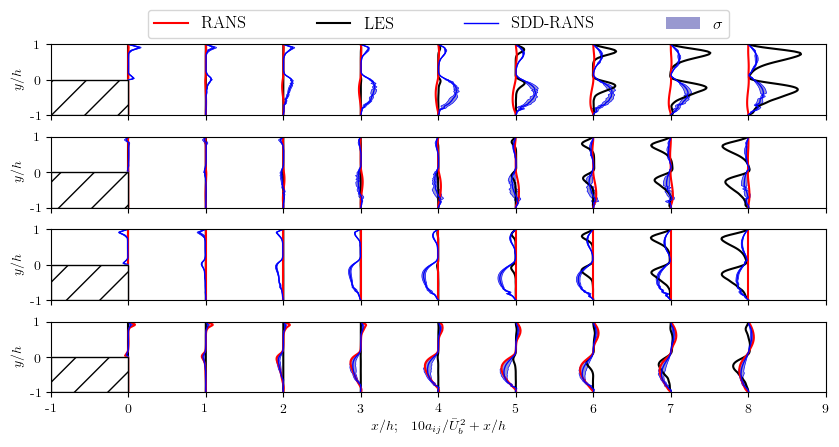}
    \caption{The anisotropic term predictions for the backwards step test flow for Reynolds number $2500$. Top to bottom: $a_{11}$, $a_{22}$, $a_{33}$ and $a_{12}$ ($a_{13}$ and $a_{23}$ are omitted due to all fields being zero).}
    \label{fig:backstepDeviatoric2500}
\end{figure}

\begin{figure}[H]
    \centering
    \includegraphics[width=0.9\textwidth]{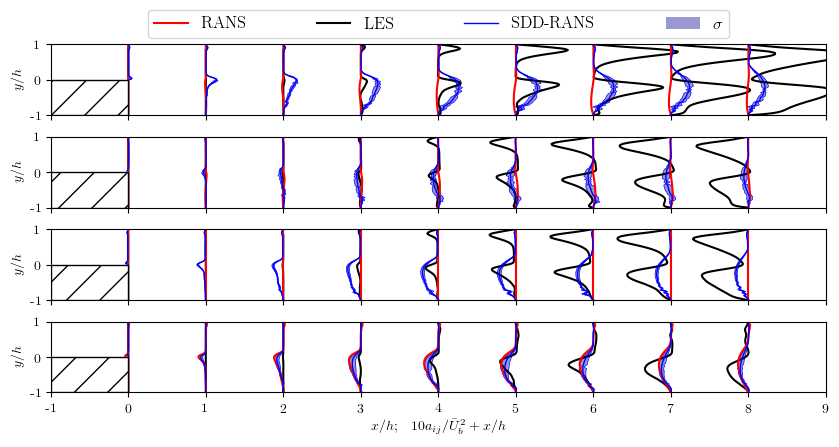}
    \caption{The anisotropic term predictions for the backwards step test flow for Reynolds number $5000$. Top to bottom: $a_{11}$, $a_{22}$, $a_{33}$ and $a_{12}$ ($a_{13}$ and $a_{23}$ are omitted due to all fields being zero).}
    \label{fig:backstepDeviatoric5000}
\end{figure}

The predicted components of the unnormalized anisotropic term $\bm{a}$, defined by $\bm{a}=k\cdot\bm{b}$ where $k$ is the baseline RANS TKE and $\bm{b}$ is the predicted anisotropic tensor, are shown in Figs.~\ref{fig:backstepDeviatoric500}-\ref{fig:backstepDeviatoric5000} for Reynolds numbers $500$, $2500$ and $5000$.
The first trend to notice is the relative consistency of SDD-RANS between all Reynolds numbers in terms of the magnitude of the mean predictions as well as variance.
This is clearly an effect of the use of training data that vary little in Reynolds number compared to the test flows.
For both $a_{11}$ and $a_{33}$ at Reynolds number $500$ and $2500$, the neural network is able to successfully predict the correct shape of the anisotropic term.
This is a notable improvement of the baseline RANS prediction which severally under-predicts all normal components.
For Reynolds number $5000$, SDD-RANS favors only a single region for $a_{11}$ and $a_{33}$ as opposed to the two regions seen in lower Reynolds numbers.
While these predictions are significant improvements over the baseline RANS solution,  there are still key discrepancies including that the anisotropic components are consistently predicted upstream compared to the LES solution.
Additionally, for terms such as $a_{11}$, the neural network under-predicts the magnitude for the larger Reynolds numbers as well as consistently under-predicts $a_{22}$.
Briefly focusing on the epistemic uncertainty of the model's predictions, the variance of the neural network's predictions are relatively small indicating   the model is over-confident for this test flow. 
Also we note that near the inlet (laminar region) there is little variance in the predicted anisotropic term.
This shows that the neural network is able to identify regions that are more uncertain than others instead of just placing a uniform variance across the entire field.

The stream-wise velocity contours of the flow for the LES, baseline RANS and the expected velocity prediction of stochastic data-driven RANS (SDD-RANS) are depicted in Figs.~\ref{fig:backstepUX500}-\ref{fig:backstepUX5000}.
To keep plot labels uncluttered, we refer to the expected values from the proposed framework as just SDD-RANS.
For each Reynolds number, the mean stream-wise velocity profiles are also illustrated with the respective predictive error bars.
We look first at the lowest Reynolds number of $500$ for which the model produced the best prediction.
Even though this corresponds to a Reynolds number furthest from the training data in Table~\ref{tab:cfdflows}, the stochastic model was able to successfully predict the appearance of the second recirculation region.
The baseline RANS simulation only predicted a single  eddy behind the step.
While the anisotropic predictions are far larger in magnitude compared to LES, the viscous forces are large enough to correct these discrepancies at lower Reynolds numbers.
We presume that these upstream over-predictions   of the anisotropic terms in magnitude   are the reason the second eddy appears closer to the inlet for SDD-RANS compared to the LES solution.
Overall, for this lower Reynolds number, the model has little variance in its predictions since the effects of the R-S prediction are dampened.

As the Reynolds number increases, the role of the R-S increases significantly as the viscous forces weaken and a very clear degradation in the predictive performance of SDD-RANS is seen.
For Reynolds number $2500$ and $5000$ (see Figs.~\ref{fig:backstepUX2500}-\ref{fig:backstepUX5000}), SDD-RANS does not yield any prediction improvement over the baseline RANS simulation with the exception of the recirculating region near the step ($x/h\leq 7.5$) where SDD-RANS is able to accurately predict the flow.
For both  higher Reynolds numbers test cases, SDD-RANS fails to predict the upper recirculation region that is present in the LES solution.
This is likely due to the under-prediction of the anisotropic components downstream seen in  Figs.~\ref{fig:backstepDeviatoric2500}-\ref{fig:backstepDeviatoric5000}.
As the R-S is increased, minor deviations in the anisotropic term are amplified resulting in potentially starkly different flow predictions~\cite{wu2018rans}.
SDD-RANS accurately captures these phenemona.
As the Reynolds number increases, so does the standard deviation indicating a loss of model confidence.
In addition, it is   clear that the model is extremely confident in its predictions towards the inlet where it is able to match the LES solution.
However, this confidence quickly diminishes downstream as flow predictions become increasingly less accurate.
With a deterministic data-driven model such indicators would not be present allowing for no interpretable information on prediction confidence without observed high-fidelity data.

\subsection{Wall Mounted Cube}
\label{subsec:wallMountedCube}
\noindent
The second test case is flow around a wall mounted cube with height $h$ as shown in Fig.~\ref{fig:wallCubeSchematic}.
Unlike the majority of the flows that have been tested by data-driven models in the literature~\cite{tracey2013application,singh2017machine,zhang2015machine,ling2016reynolds,xiao2016quantifying} as well as our training flows, this test flow contains an obstacle that is not semi-infinite.
This means that flow with this geometry cannot be modeled by a two-dimensional RANS simulation as was the case
for all previously considered flows.
Additionally, similar to the backwards step, none of our training flows contain a geometry that is similar to this.
As a result, we consider this flow an excellent test to investigate the limits of SDD-RANS in   generalizing to a true 3D test case.

The set-up of this flow consists of an uniform inlet velocity and two channel walls normal to the y-axis.
The cube is placed slightly down stream of the inlet.
The feature of interest is primarily the recirculation region behind the cube itself.
Additionally, as the Reynolds number increases, flow separation occurs on the sides of the cube.
As will be shown in the subsequent figures, this flow separation is often non-existent for the baseline RANS predictions.
While the mean flow is symmetrical about the $x-y$ plane in the middle of the channel ($z=3.5H$), we simulate the entire cube in order to observe non-symmetrical behavior in predictions.

\begin{figure}[H]
    \centering
    \includegraphics[width=1.0\textwidth]{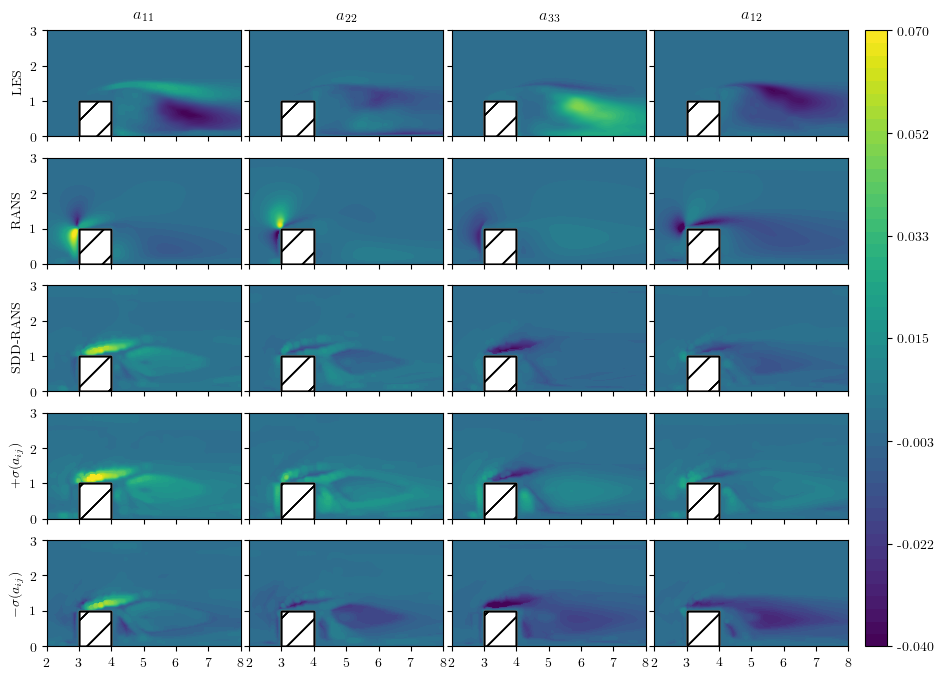}
    \caption{The anisotropic term predictions for the wall mounted cube test flow for Reynolds number $2500$ along the plane of symmetry ($a_{13}$ and $a_{23}$ are omitted due to both fields being zero on the plane of symmetry).
    From top to bottom: Time-averaged LES solution, baseline RANS prediction, SDD-RANS expected value and the predictive standard deviation error bounds.}
    \label{fig:wallRSPred2500}
\end{figure}

The prediction of the unnormalized anisotropic term along the $x-y$ plane of symmetry is shown in Fig.~\ref{fig:wallRSPred2500} for Reynolds number $2500$.
For brevity, we only show the results for a single Reynolds number since both Reynolds numbers $500$ and $5000$ lead to similar predictions.
From this figure, several positive traits are seen for both the mean neural network predictions as well as the associated uncertainty.
The bulk region shows little variability within the predictive error bounds.
Instead the variance is largely concentrated behind the obstacle.
This is a nice attribute because the baseline RANS simulation is accurate in the bulk region, thus perturbing the R-S in the bulk flow would not be of any benefit.
Another interesting feature predicted by the neural network is the concentrated region of normal stresses on top of the cube seen in  $a_{11}$ and $a_{33}$.
Similar regions, yet smaller in magnitude, appear further down-stream in the LES solution as a result of the shear layer that forms between the bulk and recirculation regions.
In addition, the SDD-RANS is able to significantly improve the R-S prediction towards the front of the cube where the baseline RANS solution is incorrect.
This includes the leading corner of the cube at $x=3$ where SDD-RANS is able to largely correct the baseline RANS solution which has sharp, unphysical predictions near the edge.

Despite the improvements to the upstream edge of the cube, the expected value of SDD-RANS has no marginal improvement on the baseline RANS prediction in the recirculation region.
However, as reflected in the predictive error bounds, the variance is larger in the recirculation zone often being able to enclose part of the true LES solution.
This is a promising result because, although the mean predictions have not improved, the model is uncertain regarding its predictions in this region.
This suggests that this area of the flow could contain physics the model has not seen before in the training data.
The stream-wise velocity contours on the plane of symmetry for the LES, baseline RANS and the expected value of SDD-RANS are depicted below in Fig.~\ref{fig:wallUXStreamCont}.
Similar to the backwards step, as the Reynolds number increases, the standard deviation of the SDD-RANS prediction also increases.
However, the magnitude of the variance is significantly smaller than that of the backwards step for higher Reynolds numbers reflecting the more accurate predictions for this problem.

\begin{figure}[H]
    \centering
    \includegraphics[width=1.0\textwidth]{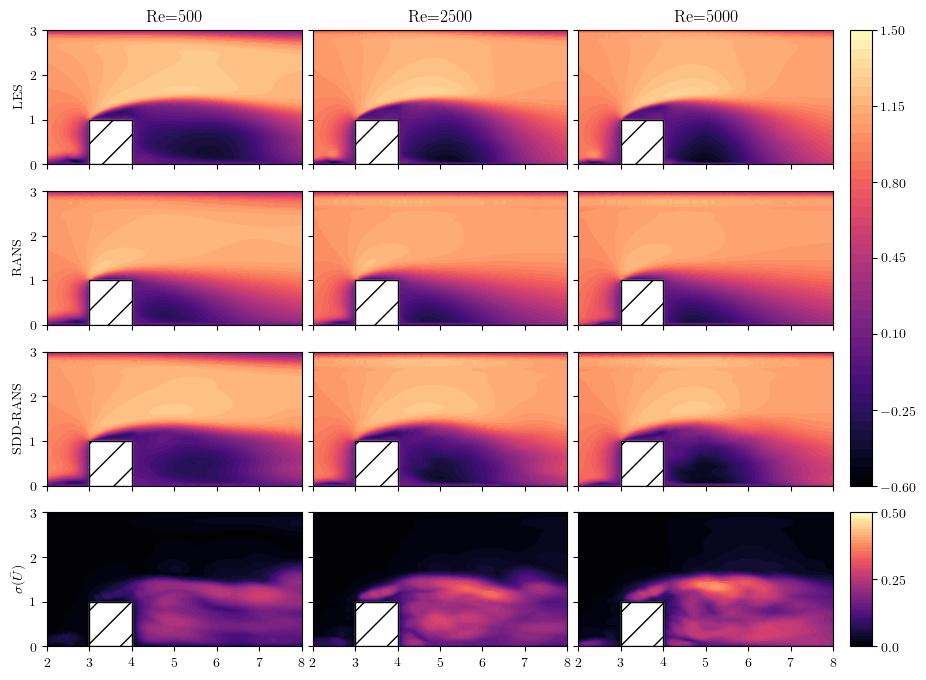}
    \caption{Normalized stream-wise mean velocity contours for Reynolds numbers $500$, $2500$ and $5000$ on the plane of symmetry. The top is the time averaged LES solution, below is the baseline RANS prediction followed by the SDD-RANS expected velocity. The fourth row shows the standard deviation field of the data-driven prediction.}
    \label{fig:wallUXStreamCont}
\end{figure}

\begin{figure}[H]
    \centering
    \includegraphics[width=0.48\textwidth]{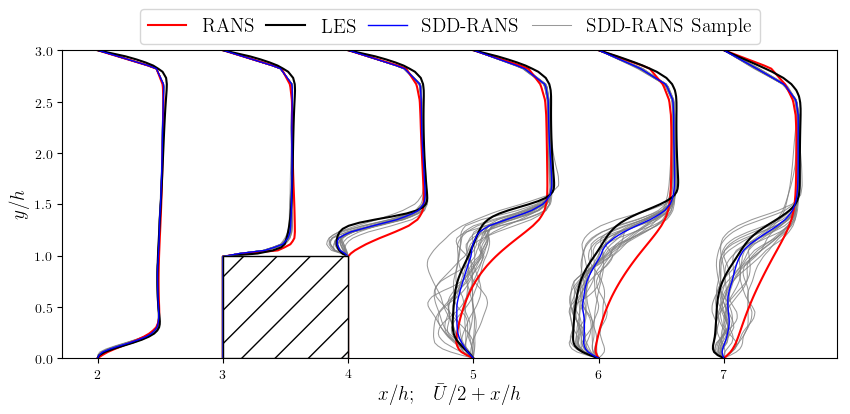}
    ~
    \includegraphics[width=0.48\textwidth]{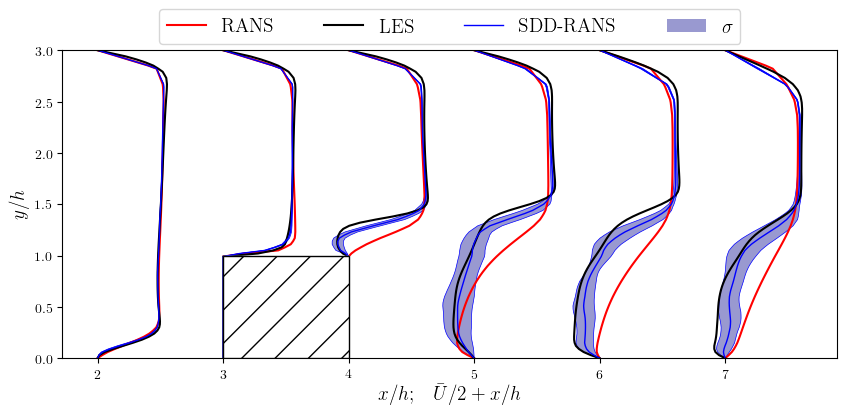}
    \caption{Normalized stream-wise velocity profiles for the baseline RANS, high-fidelity LES, and SDD-RANS predictions on the plane of symmetry at six different locations in the stream-wise direction for Re$=500$.
    The left shows the SDD-RANS velocity samples, and the right shows the respective predictive error bars for each profile.}
    \label{fig:wallUXStreamProfile500}
\end{figure}

\begin{figure}[H]
    \centering
    \includegraphics[width=0.48\textwidth]{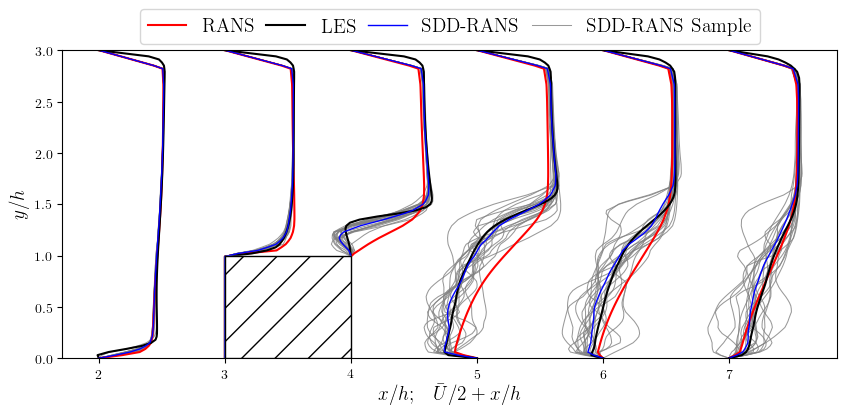}
    ~
    \includegraphics[width=0.48\textwidth]{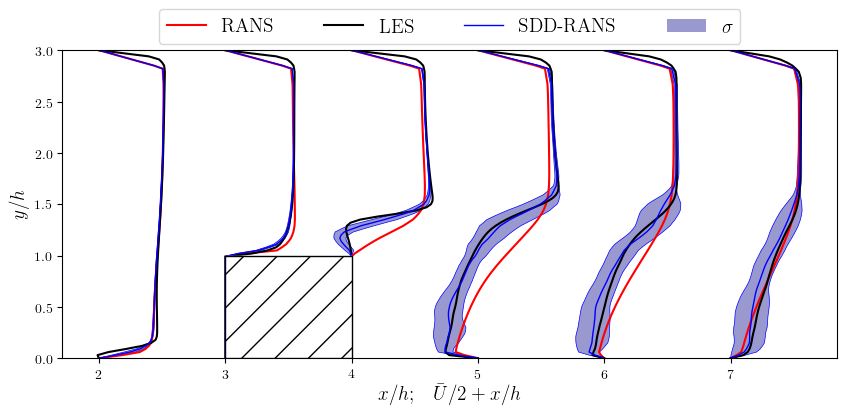}
    \caption{Normalized stream-wise velocity profiles for the baseline RANS, high-fidelity LES, and SDD-RANS predictions on the plane of symmetry at six different locations in the stream-wise direction for Re=5000.
    The left shows the SDD-RANS velocity samples, and the right shows the respective predictive 
    error bars for each profile.}
    \label{fig:wallUXStreamProfile5000}
\end{figure}

To take a closer look at the performance of SDD-RANS, stream-wise velocity profiles are plotted for both Reynolds number $500$ and $5000$ in Figs.~\ref{fig:wallUXStreamProfile500} and~\ref{fig:wallUXStreamProfile5000}, respectively.
For each, a plot containing the resulting SDD-RANS velocity field samples are shown as well as the predictive standard deviation error bars.
As expected the variance in the velocity samples increases with the increased Reynolds number, however this only occurs in the recirculation region where the instantaneous flow is turbulent.
In the bulk region above the recirculation zone, the variance remains small.

A significant improvement by SDD-RANS for both Reynolds numbers is the prediction of the detached flow on top of the cube ($x/h=4$) which the baseline RANS fails to capture.
This is largely due to the large increase of normal stresses from the neural network predictions around the surrounding walls of the cube which results in the shear layer forming above the obstacle.
For both Reynolds numbers cases, improvements in the recirculation region prediction are present. However,  for the lower Reynolds number case SDD-RANS leads to more accurate predictions.

\begin{figure}[H]
    \centering
    \includegraphics[width=1.0\textwidth]{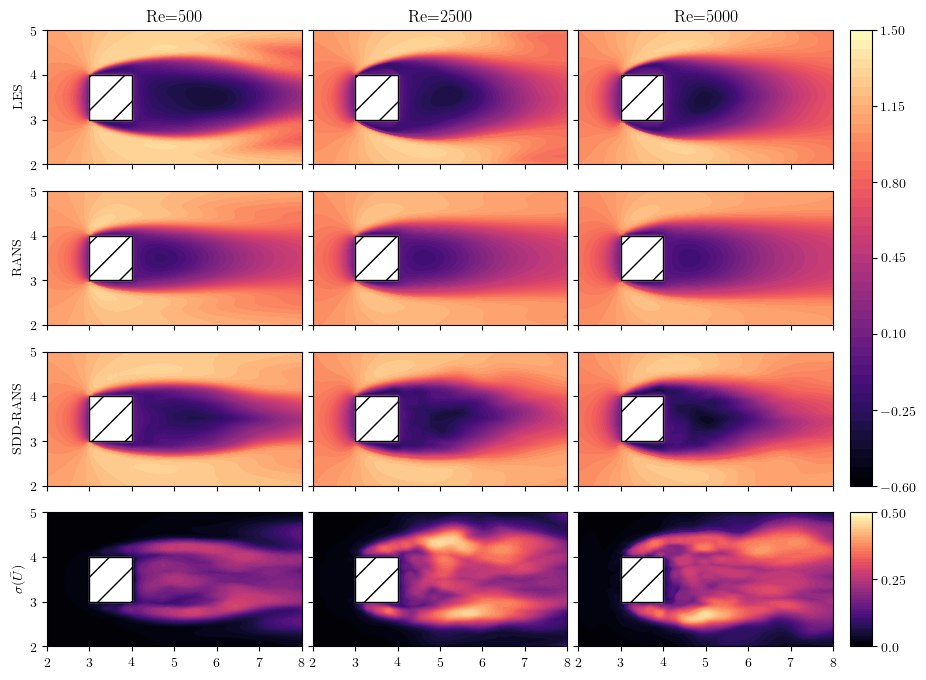}
    \caption{Normalized stream-wise mean velocity contours for Reynolds numbers $500$, $2500$ and $5000$ on the $y=0.5h$ plane. The top is the time averaged LES solution, below is the baseline RANS prediction followed by the SDD-RANS expected velocity. The fourth row shows the standard deviation field of the data-driven prediction.}
    \label{fig:wallUXSpanCont}
\end{figure}

\begin{figure}[H]
    \centering
    \includegraphics[width=0.48\textwidth]{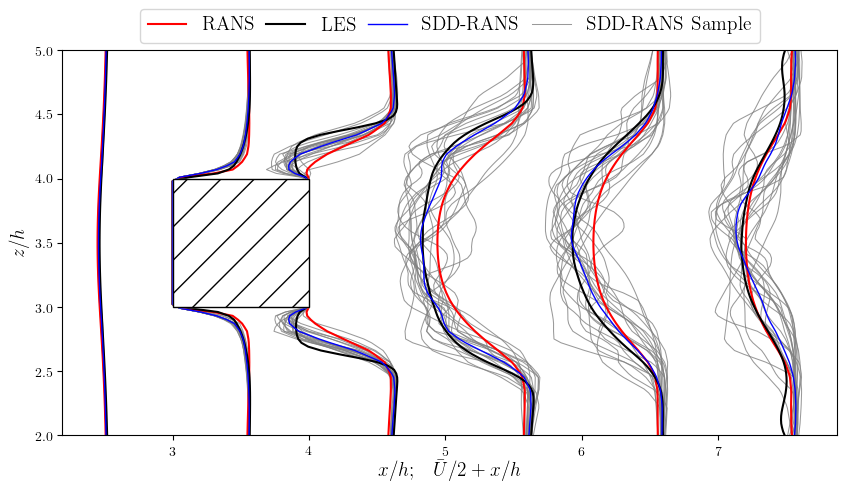}
    ~
    \includegraphics[width=0.48\textwidth]{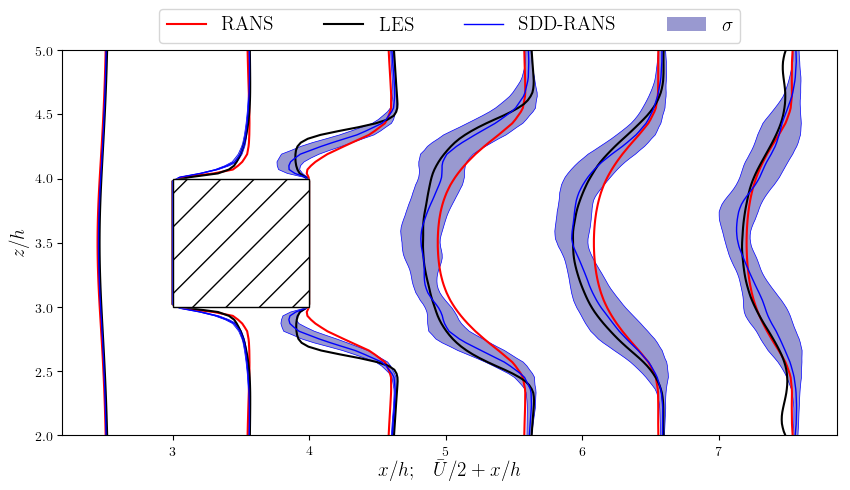}
    \caption{Normalized stream-wise velocity on the $y=0.5h$ plane for the baseline RANS, high-fidelity LES, and SDD-RANS predictions on the plane of symmetry at six different locations in the stream-wise direction for Re$=2500$.
    The left shows the SDD-RANS velocity samples, and the right shows the respective predictive error bars for each profile.}
    \label{fig:wallUXSpanProfile2500}
\end{figure}

Since this obstacle is not semi-infinite, the velocity contours for the horizontal plane at $y=0.5h$ are shown in Fig.~\ref{fig:wallUXSpanCont}.
Similarly, velocity profiles are plotted for Reynolds number $2500$ in Fig.~\ref{fig:wallUXSpanProfile2500}.
In general, we can see the same trends as previous results for which the variance increases with Reynolds number.
We note that SDD-RANS is able to predict the presence of the detached flow on the side of the cube.
While some asymmetry exists in the SDD-RANS predictions, the predictions overall  retain a general symmetric profile.
It is likely that increasing the number of samples of R-S fields would further improve the symmetry and smoothness of predictions.

\begin{figure}[H]
    \centering
    \includegraphics[width=0.7\textwidth]{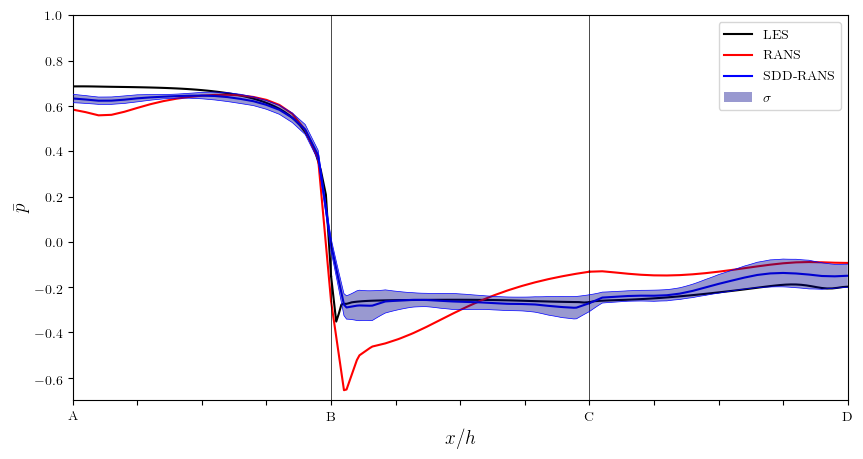}
    ~
    \includegraphics[width=0.27\textwidth]{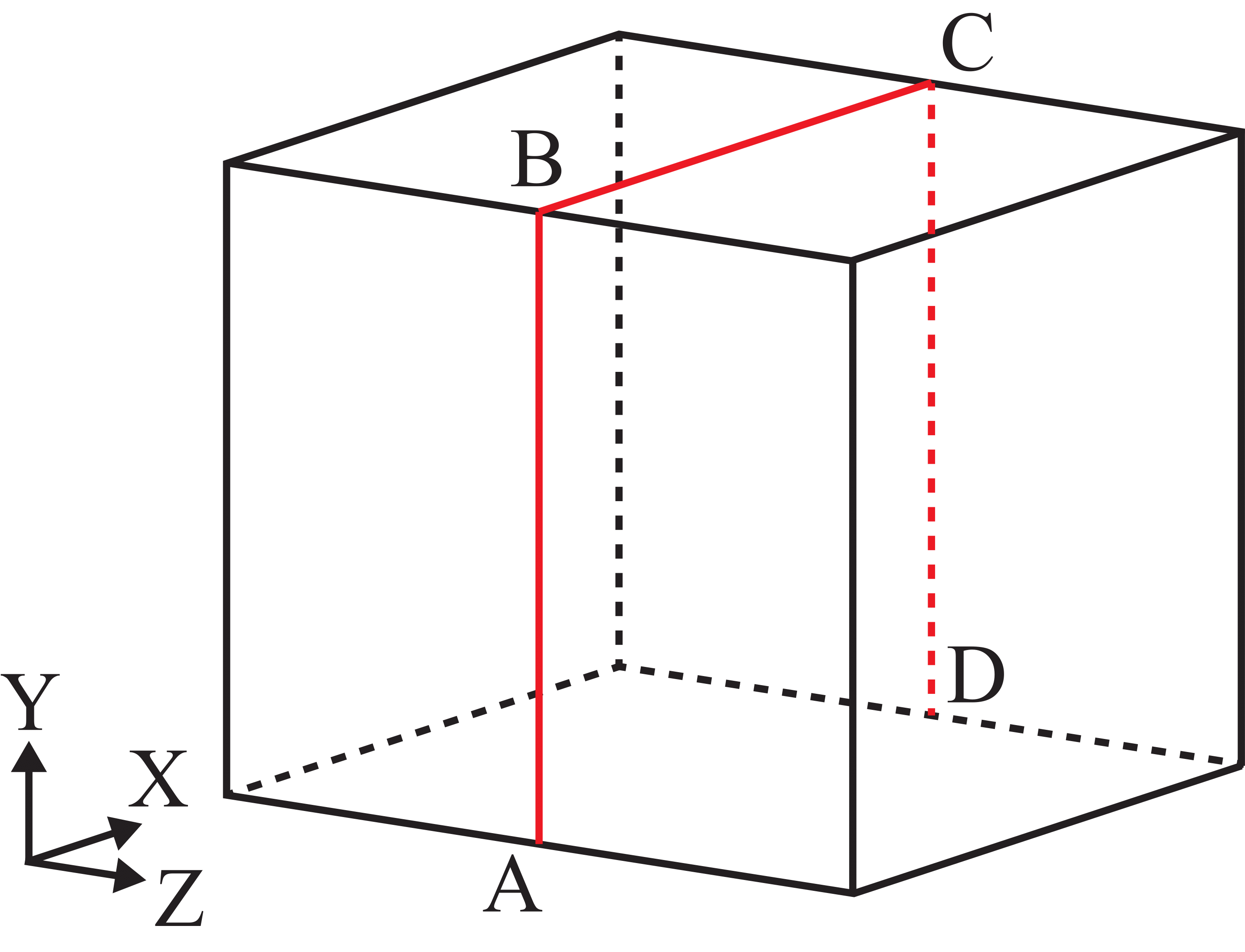}
    \caption{Normalized mean surface pressure profile on the plane of symmetry ($z=3.5h$) for Re$=5000$.}
    \label{fig:wallCubePressure1}
\end{figure}

\begin{figure}[H]
    \centering
    \includegraphics[width=0.7\textwidth]{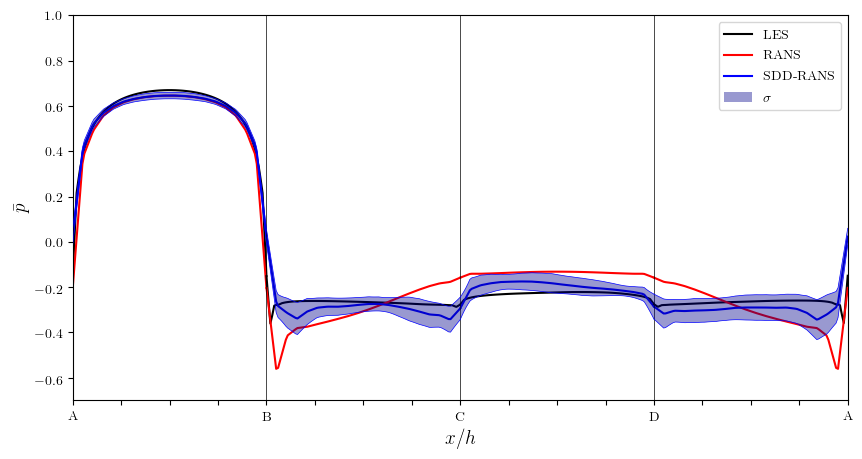}
    ~
    \includegraphics[width=0.27\textwidth]{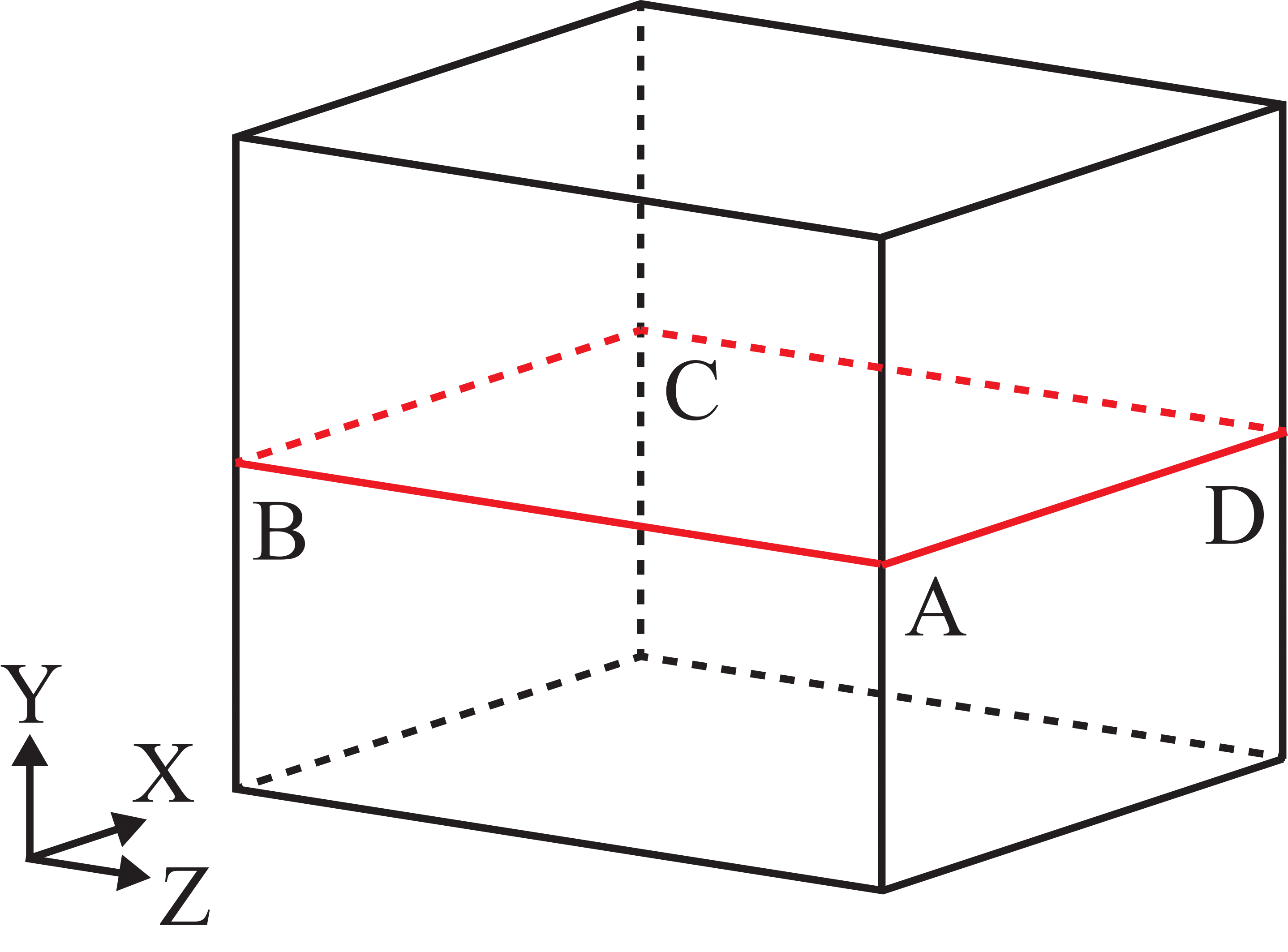}
    \caption{Normalized mean surface pressure profile on the plane $y=0.5h$ for Re$=5000$.}
    \label{fig:wallCubePressure2}
\end{figure}

This framework allows probabilistic bounds to be calculated for other fluid properties such as pressure, drag, shear stress, etc.
For example, two pressure profiles along the face of the wall mounted cube are plotted   in Figs.~\ref{fig:wallCubePressure1} and~\ref{fig:wallCubePressure2}.
In general we see that SDD-RANS is able to provide an improved prediction compared to the baseline RANS.
Similar to the anisotropic components, SDD-RANS corrects the unphysical pressure drop that occurs on the edge of the leading cube face in the baseline RANS simulation.
The uncertainty for the predictive pressure is also very reasonable nearly capturing the true LES prediction for all faces.
Similar to the velocity predictions in Figs.~\ref{fig:wallUXStreamCont} and \ref{fig:wallUXSpanCont}, the variance of the pressure on the upstream face of the cube is significantly smaller reflecting the model's confidence in this laminar region.

\section{Conclusions}
\label{sec:Conclusions} 
\noindent
As the CFD community continues to investigate the use of machine learning tools for data-driven modeling, the need to accurately quantify the induced uncertainties from the use of such models becomes  essential.
In this work, we have presented a novel framework that allows for the quantification of such model form uncertainty.
To satisfy invariant properties, we use the neural network architecture originally proposed by Ling~\etal~\cite{ling2016reynolds}.
Using Stein variational gradient decent and following the work of Zhu and Zabaras~\cite{zhu2018bayesian}, we extended this invariant neural network model to a Bayesian deep neural network to allow us to compute the distribution of the anisotropic R-S tensor for a given baseline solution. 
To   propagate the  uncertainty of this model to fluid flow quantities of interest,   a stochastic data-driven RANS algorithm is proposed that utilizes standard Monte Carlo simulation.
The integrated framework was  rigorously investigated on two flows to observe its generalization property. 

In the presented implementation of this framework, we found that the invariant neural network used to model the anisotropic tensor proved difficult to train and yield satisfactory predictions for unseen flows and geometries.
From our studies, we hypothesize that using just the five invariant inputs does not provide enough descriptive information to accurately map from the coarse to high-fidelity flow physics.
Although the network contains desired invariant properties, other flow quantities would most likely need to be used as model inputs to improve the quality of predictions. 
Thus a critical area to be investigated is the development of more accurate Reynolds stress representations by identifying the important local- and non-local variables that influence its values.
The potential use of spatial correlations and   information at neighboring nodes (non-local models) may prove to be extremely beneficial.
Such an approach, while difficult to implement for non-uniform grids, can be easily applied in the context of   convolutional neural networks that are capable of mapping many high-dimensional inputs to multi-outputs of high-dimensionality.
While a number of other models in the literature can yield much better training predictions, most of these models remain only useful to a small family of flows resembling those in the training dataset. 
The generalization property of these models remains an open   problem for the data-driven community.

With improvements in the representation of the tuned Reynolds stress in the RANS equations, we believe that this framework can provide extremely beneficial information for data-driven models.
While the most obvious application is its use to assess a given model's predictive confidence, the developed framework can also be used to identify locations of potentially lower accuracy.
This could be useful for identifying areas that may require finer mesh resolutions or high-fidelity simulations.
Additionally, the use of a Bayesian neural network allows us to compute predictive bounds for the quantities of interest.
This can be extremely useful in cases where the training data is limited.
Even though the use of the Bayesian neural network and   SDD-RANS requires more computational time than  deterministic data-driven approaches, we found that when compared to high-fidelity simulations the computational cost remains low.

The use of a Bayesian neural network opens up the potential of implementing experimental design techniques by investigating the impact of training data on the quality of the model's predictions.
This can range from the assessment of a limited data case or how specific training flows at various Reynolds numbers impact a specific test case prediction.
Finally, a detailed analysis of epistemic uncertainty would be  beneficial to the data-driven turbulence modeling community.

\section*{Acknowledgements}
\noindent
The authors acknowledge support  from the Defense Advanced Research Projects Agency (DARPA) under the Physics of Artificial Intelligence (PAI) program (grant No. HR00111890034). 
The work of NG is also supported by a National Science Foundation (NSF) Graduate Research Fellowship Program grant No. DGE-$1313583$. 
The computing was facilitated by the resources of the  NSF supported ``Extreme Science and Engineering Discovery Environment'' (XSEDE) on the Bridges and Bridges-GPU cluster through the startup allocation No. TG-CTS$180011$ and research allocation No. TG-CTS$180038$.
Additional computing resources were provided by the University of Notre Dame's Center for Research Computing (CRC). 

\section*{References}

\bibliography{mybibfile}

\end{document}